\documentclass[bibyear]{aa}
\usepackage{times}
\usepackage{amsmath,mathrsfs,amssymb,amsbsy}
\usepackage{graphicx}
\usepackage{subfigure}
\usepackage{paralist}
\usepackage{color}
\usepackage{natbib}
\usepackage{aas_macros}
\usepackage{float}
\usepackage[colorlinks=true,linkcolor=blue, citecolor=blue]{hyperref}%
\graphicspath{{plots/}}
\usepackage{grffile} % In order to find multidot graphic file extensions

\newcommand{\kms}{\,km\,s$^{-1}$} % kilometres per second
\usepackage{gensymb} % allows to use \degree and obtain the degree symbol
\usepackage{soul} % To bar text

\newcommand{\teff}{$T_{\rm eff}$}
\newcommand{\logg}{$\log g$}
\newcommand{\meta}{\rm{[M/H]}}

\newcommand{\Mini}{\rm {M_{ini}} } 

\newcommand{\cgs}{\,{$\rm cm~s^{-2}$}}

\def\afe{\rm [\alpha/Fe]}

\def\kms{\,{\rm km~s^{-1}}}
\def\kpc{\,{\rm kpc}}

\def\dex{\,{\rm dex}}
\def\Gyr{\,{\rm Gyr}}
\def\vphi{V_\phi}
\def\vz{V_Z}
\def\vr{V_R}

\def\ltsima{$\; \buildrel < \over \sim \;$}
\def\simlt{\lower.5ex\hbox{\ltsima}}
\def\gtsima{$\; \buildrel > \over \sim \;$}
\def\simgt{\lower.5ex\hbox{\gtsima}}

\newcommand{\gaiaG}{$G$}
\newcommand{\BP}{$G_{{BP}}$}
\newcommand{\RP}{$G_{{RP}}$}
\newcommand{\EBPRP}{E(G_{BP}-G_{RP})} % will have to put $$. 
\newcommand{\gspspec}{GSP-Spec}
\newcommand{\gspphot}{GSP-Phot}

\def\ltsima{$\; \buildrel < \over \sim \;$}
\def\simlt{\lower.5ex\hbox{\ltsima}}
\def\gtsima{$\; \buildrel > \over \sim \;$}
\def\simgt{\lower.5ex\hbox{\gtsima}}

\providecommand{\orcit}[1]{\protect\href{https://orcid.org/#1}{\protect\includegraphics[width=8pt]{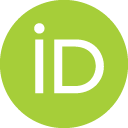}}}
    
\begin{document}

\title{Stellar ages, masses, extinctions and orbital parameters based on spectroscopic parameters of Gaia DR3\thanks{The parameters computed in this paper have been prepared in the context of Gaia's Performance verification paper concerning the Chemical cartography of the Milky Way \citep{GaiaDR3_PVP_chemicalCartography}. }}
\author{
	G.~Kordopatis\orcit{0000-0002-9035-3920},\inst{\ref{oca}}\thanks{\email{georges.kordopatis@oca.eu} }
	M.~Schultheis\orcit{0000-0002-6590-1657}\inst{\ref{oca}},
	P.~J.~McMillan\orcit{0000-0002-8861-2620}\inst{\ref{lund}},
	%(not final)
	P.~A.~Palicio\orcit{0000-0002-7432-8709}\inst{\ref{oca}},
	P.~de~Laverny\orcit{0000-0002-2817-4104}\inst{\ref{oca}},
	A.~Recio-Blanco\orcit{0000-0002-6550-7377}\inst{\ref{oca}},
	O.~Creevey\orcit{0000-0003-1853-6631}\inst{\ref{oca}},
	M.~A.~\'Alvarez\orcit{0000-0002-6786-2620}\inst{\ref{Coruna}}, 
	R.~Andrae\orcit{0000-0001-8006-6365}\inst{\ref{mpia}},
	%M.~Fouesneau\orcit{0000-0001-9256-5516}\inst{\ref{mpia}},
	E.~Poggio\orcit{0000-0003-3793-8505}\inst{\ref{oca}, \ref{torino}}, 
	E.~Spitoni\orcit{0000-0001-9715-5727}\inst{\ref{oca}, \ref{aarhus}},
	G.~Contursi\orcit{0000-0001-5370-1511}\inst{\ref{oca}},
	H.~Zhao\orcit{ 0000-0003-2645-6869}\inst{\ref{oca}},
	I.~Oreshina-Slezak\inst{\ref{oca}},
	C.~Ordenovic\orcit{0000-0003-0256-6596}\inst{\ref{oca}}, 
	A.~Bijaoui\inst{\ref{oca}} 
 	%Others,\inst{\ref{oca}, \ref{lund}}
	 % 
	}
\institute{
	Universit\'e C\^ote d'Azur, Observatoire de la C\^ote d'Azur, CNRS, Laboratoire Lagrange, Nice, France\label{oca}
	\and
	Lund Observatory, Lund University, Department of Astronomy and Theoretical Physics, Box 43, SE-22100, Lund, Sweden\label{lund}
	\and
	CIGUS CITIC - Department of Computer Science and Information Technologies, University of A Coru\~{n}a, Campus de Elvi\~{n}a s/n, A Coru\~{n}a, 15071, Spain\label{Coruna}
	\and
	Max Planck Institute for Astronomy, K\"{ o}nigstuhl 17, 69117 Heidelberg, Germany\label{mpia}
	\and
	Osservatorio Astrofisico di Torino, Istituto Nazionale di Astrofisica (INAF), I-10025 Pino Torinese, Italy\label{torino}
	\and
	Stellar Astrophysics Centre, Department of Physics and  Astronomy, Aarhus University, Ny Munkegade 120, DK-8000 Aarhus C, Denmark \label{aarhus}
	}
\abstract
%Context
{Gaia's third data release provides radial velocities for 33 million stars and spectroscopically derived atmospheric parameters 
for more than five million targets. When combined with the astrometric data, these allow us to derive 
orbital and stellar parameters that are key in order to understand the stellar populations of the Milky Way and perform galactic archaeology.  }
 %Aims
 {We use the calibrated atmospheric parameters, 2MASS and Gaia-EDR3 photometry, and  parallax-based distances to compute the ages, initial stellar masses and reddenings for the stars with spectroscopic parameters. We also derive the orbits for all of the stars with measured radial velocities and astrometry, adopting two sets of line-of-sight distances from the literature.}
%Methods
{Four different sets of ages, masses and absolute magnitudes in different photometric bands are obtained through an isochrone fitting method, considering different  combinations of  input parameters. The reddenings are obtained by comparing the  observed colours with the ones obtained from the isochrone projection. Finally, the orbits are computed adopting an axisymmetric potential of the Galaxy. }% and commonly used orbital derivation methods and codes. }
%Results
{ 
Comparisons with reference catalogues of field and cluster stars suggest that reliable ages are obtained for stars younger than 9-10\Gyr~when the estimated relative age uncertainty is $<50$ per cent. For older stars,  ages tend to be under-estimated.
The most reliable stellar type for age determination are  turn-off stars, even when the input atmospheric parameters have large uncertainties. Ages for giants and main-sequence stars are retrieved with uncertainties of the order of $2\Gyr$ when extinction towards the star's line-of sight  is smaller than $A_V\lesssim 2.5$\,mag.
} 
%Conclusions
{The catalogue of ages, initial stellar masses, reddenings, Galactocentric positions and velocities, as well as the stellar orbital actions, eccentricities, apocentre, pericentre and maximum distance from the Galactic plane reached during their orbits,  is made publicly available to be downloaded. With this catalogue, the full chemo-dynamical properties of the extended Solar neighbourhood unfold, and allow us to better identify the properties of the spiral arms, to parameterise the dynamical heating of the disc, or to thoroughly study the  chemical enrichment of the Milky Way. }


\titlerunning{GDR3 catalogue of ages and orbits}
\authorrunning{G.~Kordopatis et al.}

\maketitle

%%%%%%%%%%%%%%%%%%%%%%%%%%%%
\section{Introduction}
\label{sec:Introduction}
Galactic archaeology  relies on the fact that stellar fossil records (chemical abundances and orbital properties) can be used to rewind time and unravel the history of the Milky Way \citep{Freeman02}.  However, in order to put the past events of our Galaxy into perspective, one also needs to have access to the stellar ages. Yet, stellar ages are intrinsically difficult to obtain for a large amount of field stars \citep[][and references therein]{Soderblom10}. 
 Asteroseismology, i.e. the analysis of the oscillating frequencies of stars, offers nowadays one of the most precise ways to determine the stellar ages, however, it is limited to relatively bright  (and nearby) targets \citep[e.g.][]{Miglio13, Pinsonneault18, Stello22}. 
When no asteroseismic information is available, both precise and accurate atmospheric parameters (effective temperature, \teff, surface gravity, \logg, overall metallicity, \meta), and/or de-reddened colours and magnitudes, are required in order to find  the best fitting model from a set of theoretical isochrones.  This tabulated synthetic star will hence allow us to derive the set of parameters that are not direct observables, such as the age, the mass and the absolute magnitudes in different photometric bands.  This technique is the so-called isochrone projection method \citep[e.g.][]{Jorgensen05}. 

The advent of the large stellar spectroscopic surveys, more than fifteen years ago, saw the development of the first isochrone projection codes that had as a main goal deriving the so-called spectroscopic parallaxes, i.e. the line-of-sight distances via the use of the distance modulus and the (projected) absolute magnitudes. Although these methods delivered naturally the stellar ages as an output \citep[e.g.][]{Zwitter10, Binney14a}, the uncertainties that were associated to them were beyond the acceptable for Galactic archaeology, unless stellar distance was known in order to constrain the projection \citep[e.g.][ for an application with stars in the Carina dwarf spheroidal galaxy]{Kordopatis16a}.  

Data gathered by the European Space Agency satellite Gaia  \citep{Gaia, GaiaDR1, GaiaDR2, GaiaEDR3, GaiaDR3} has undoubtedly opened a new era regarding the age determination. Stellar parallaxes ($\varpi$), and heliocentric line-of-sight distances derived from them \citep[e.g.][]{Bailer-Jones15, Luri18, Schonrich19, Anders19}, can now be used in the projection, in combination with the precise photometry (\gaiaG, \BP, \RP), reducing significantly the age uncertainties \citep[e.g.][]{McMillan18, Sanders18, Queiroz18, Leung19, Feuillet19}.

Within Gaia's Data Processing Analysis Consortium (DPAC), the Astrophysical parameters inference system (Apsis)  \citep{Bailer-Jones13, GaiaDR3-CU8-1, GaiaDR3-CU8-2, GaiaDR3-CU8-3} is the work chain in charge of obtaining the physical parameters\footnote{As opposed to the astrometric parameters.} of the  targets observed by the satellite. More specifically, the stellar parameters for single stars are obtained from six different modules, reflecting either the different nature of the input data taken into account by each of them, and/or the different stellar types they are dealing with. Amongst these modules, 
the  General Stellar Parametrizer from Photometry (\gspphot), the General Stellar Parametrizer from Spectroscopy (\gspspec) and the Final Luminosity Age Mass Estimator (FLAME)  are the ones deriving parameters for most of the targets. 

On the one hand, \gspphot\ obtained the \teff, \logg, \meta, absolute magnitude in the $G$ band, radius, extinction, and line-of-sight distance of $\sim400$ million stars of OBAFGKM spectral-type, based on the analysis of the very low-resolution Blue and Red spectrophotometers (120 flux-points, each, with a resolving power ranging from 20 to 90 depending on the wavelength), in combination with the parallaxes. Overall, parameters show median absolute deviation (MAD)  compared to the literature of $\sim 220$\,K, 0.25\dex, 0.26\dex\ for \teff, \logg, \meta,  respectively. 
% The above numbers are the average of Tables 3.18, 3.19 and 3.20 of the online doc. 
There are limitations, however, depending on the stellar type, the metallicity regime and the true extinction along the line-of-sight, due to the low resolution of the input data, and to the degeneracy between \teff\ and extinction \citep[][]{Andrae18, GaiaDR3-GSPphot}
 % GSPphot MADs are the ones from the online documentation, based on an average of the Tables of GSPphot. 

On the other hand, \gspspec\ avoids the limitations of \gspphot\ by analysing the medium-resolution spectra ($R=\lambda/\delta \lambda\approx11\,000$) gathered by the Radial Velocity Spectrometer (RVS). However, only a fraction of stars ($\sim 5\cdot10^6$ targets of FGK stellar-type) have published parameters, i.e. the targets whose spectra have an adequate signal-to-noise (S/N$>$20, \citealt{GaiaDR3-GSPspec}). The stated MAD compared to literature values for \teff, \logg\ and \meta\ are $61$\,K, $0.14$\dex\ and $0.09$\dex, 
% The above numbers are the sigma_mad of Fig. 10 of the GSPspec divided by 1.48 to make it MAD.
whereas $\alpha$-elements  enhancement ($\afe$) and a handful of elemental abundances are also provided for a subsample of them.

Finally, the FLAME module derived two sets of masses ($\mathcal{M}$), ages ($\tau$) and evolutionary phases  for most Gaia sources, based on either  \gspphot\ or \gspspec\ results. These evolutionnary parameters were obtained by first computing a bolometric correction for each target (based on the stars' effective temperature, surface gravity and metallicity), then deriving the bolometric luminosity $\mathcal{L}$ (via the relation linking  $\mathcal{L}$ with the absolute magnitude of the star), and finally getting the stellar radius $\mathcal{R}$ via the Stefan-Boltzmann relation linking $\mathcal{R}$ with $\mathcal{L}$. The luminosities and radii are then projected on BASTI \citep{Hidalgo18} isochrones of solar metallicity to obtain $\tau$ and $\mathcal{M}$.

In this paper, we build upon the Gaia public catalogues, and  focus on the  RVS sample of GDR3. 
This sample  contains $\sim 33 \cdot 10^6$ stars with radial velocity measurements and $~5 \cdot10^6$ stars with spectroscopic parameters and abundances \citep{GaiaDR3-GSPspec, GaiaDR3-CU8-2}. We perform an improved projection on the isochrones that takes into account the necessary \gspspec~ calibrations of metallicity and surface gravities\footnote{Due to the tight schedule of DPAC to publish GDR3,  FLAME parameters based on \gspspec's parameters did not use the calibrated values, as the latter came after the computation of parameters from the former. },  stellar distance (in a different way than FLAME does, using the \citealt{Bailer-Jones21} values), Infra-red  \citep[2MASS, ][]{Skrutskie06}, and EDR3 photometry. 
Furthermore, for the community's convenience, we also compute and provide the catalogue of the 3-dimensional galactocentric positions, (cylindrical) velocities and orbital parameters of all of the GDR3 RVS stars using an updated version of the axisymmetric Galactic potential of \citet{McMillan17} and the Galpy code \citep{Bovy15}.

Section~\ref{sec:isochrone_section} describes the way the ages, the masses, the absolute magnitudes and the reddenings have been computed. In particular, this section contains a description of the isochrone sets that we adopted, and the validation of the projected parameters.  
Section~\ref{sec:positions_orbits_illustrations} discusses how the orbital parameters have been obtained and illustrates how the ages and masses correlate with them. 
Finally, Sect.~\ref{sec:conclusions} concludes. The catalogue containing all of these parameters can be downloaded on the following address: \url{https://ftp.oca.eu/pub/gkordo/GDR3/}, with the different columns described in Table~\ref{tab:catalogue}.

\section{Catalogue of ages, initial stellar masses, extinctions and reddenings}
\label{sec:isochrone_section}

This section first describes the sets of isochrones we employ in this work (Sect.~\ref{sec:isochrones_set}). The  mathematical basis of the algorithm is presented in Sect.~\ref{sec:method}, and the priors that are used to optimise the isochrone projection are justified in Sect.~\ref{sec:age_prior}.  The method's accuracy and precision  is tested on synthetic data extracted from isochrones in Sect.~\ref{sec:synthetic_tests}. Points to consider, when projecting real data, such as colour-\teff~ calibration or the effect of $\alpha$-enhancement on the overall metallicity are discussed in Sect.~\ref{sec:real_life}.  
The actual projection of the calibrated \gspspec~parameters is done in Sect.~\ref{sec:dataset_definitions}. The compiled age and mass sample is compared and validated with respect to reference catalogue of field and cluster stars in Sect.~\ref{sec:age_mass_validation}.  Finally, reddenings and absolute magnitudes derived from the projection are discussed in Sect.~\ref{sec:extinctions} and \ref{sec:Absmags}.

%%%%%%%%%%%%%%%%%%%%%%%%%%%%%%%%%%
\subsection{Definition of the isochrone set}
\label{sec:isochrones_set}

\begin{figure*}[ht!]
\begin{center}
\includegraphics[width=\linewidth, angle=0]{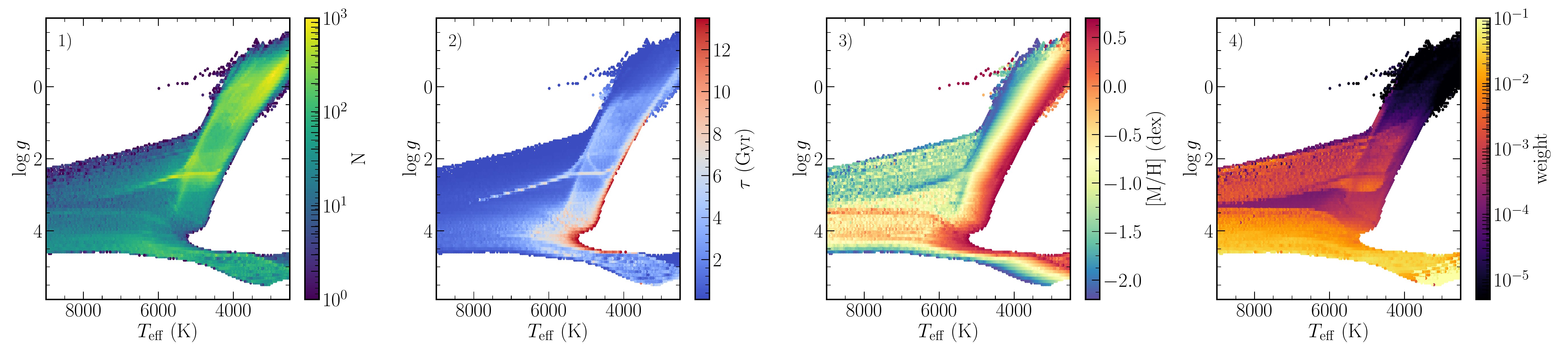}\\
\includegraphics[width=\linewidth, angle=0]{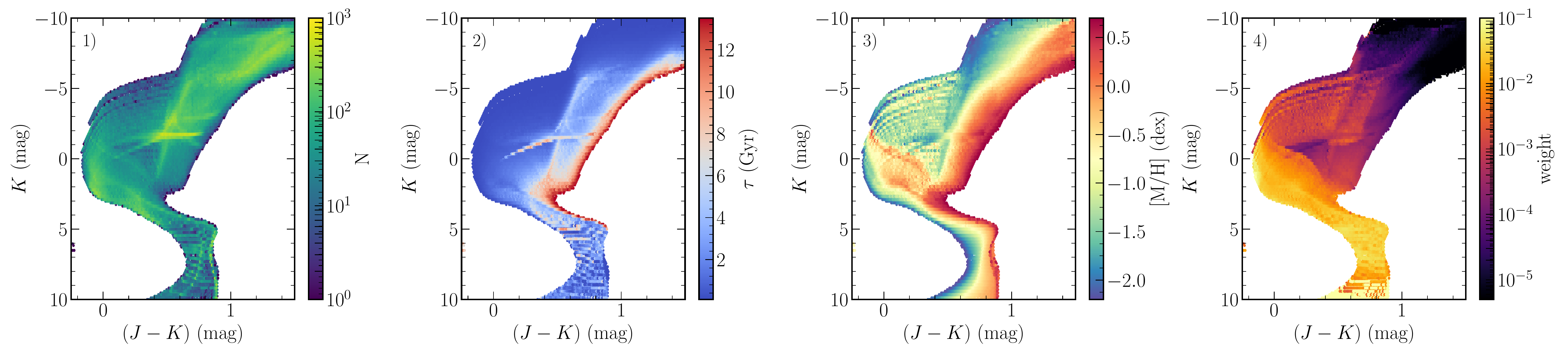}\\
\caption{\teff\,$vs$ \logg~ (top) and $(J-K_s)$ $vs$ $K_s$~(bottom) hexbin plots  of the isochrones employed in this work to derive the ages, masses, and reddenings. Panels 1) are colour-coded by density of points, panels 2) by average age, panels 3) by average metallicity and panels 4) by average evolutionary weight (factor $\beta_i$ in Eq.~\ref{eq:isochrone_weights}). }
\label{fig:isochrones}
\end{center}
\end{figure*}

We used the PARSEC stellar tracks \citep{Bressan12} version 1.2S \citep[][]{Tang14, Chen15} up to the beginning of Asymptotic Giant Branch (AGB) and the COLIBRI $S_{37}$ tracks up to the end of the AGB phase \citep[][]{Pastorelli20, Marigo13, Rosenfield16}, in combination with the online interpolator tool\footnote{version 3.5, \url{http://stev.oapd.inaf.it/cgi-bin/cmd_3.5}} of the Padova group, to compute a library of isochrones spanning the ages between $0.1$ and $13.5\Gyr$, logarithmically spaced by a $\log(\tau)$ step of $0.05$, and the metallicities between $-2.2$ and $+0.7\dex$ with a step of $0.05\dex$ ($i.e.$ of the same order as \gspspec's uncertainties on metallicity). 
The metallicity values for the isochrones are obtained using the approximation $\meta=\log(Z/X)-\log(Z/X)_\odot$, with $(Z/X)_\odot=0.0207$  and $Y=0.2485+1.78Z$, and where $X,Y,Z$ are, respectively,  the  hydrogen, helium and metal abundances, by mass \citep{Caffau11, Basu97}.

In total, 2\,301 different isochrone tracks, containing 842\,494
tabulated \teff, \logg, \meta, $\tau$, initial stellar mass ($M_{\rm ini}$), absolute magnitudes in the \gaiaG, \BP, \RP~EGDR3 photometric bands and in the 2MASS $J$, $H$, $K_s$ bands were obtained in such a way.  Their \teff~vs~\logg~and $(J-K_s)$ vs $K_s$ distribution and properties are shown in Fig.~\ref{fig:isochrones}. This set of isochrones is the one that is going to be used in what follows in order to project the atmospheric parameters on.

\subsection{Isochrone projection algorithm}
\label{sec:method}

The method described below was initially inspired by \citet{Zwitter10} and early implementations of it have already been successfully employed for either distance computation \citep{Kordopatis11b, Kordopatis13a, Kordopatis15b, Recio-Blanco14},  age derivation \citep{Kordopatis16a, Magrini17, Magrini18, Santos-Peral21} or reddening estimation \citep{Schultheis15, Zhao21}. For completeness, we summarise in what follows the method, which contains specific changes that have been implemented for the  Gaia-RVS application.

We consider a star in the Gaia catalogue, to which is associated a set of observed parameters $\hat \theta_k$ ($k\equiv$ \teff, \logg, \meta, \gaiaG, $J$, $H$, $K_s$,...)  and uncertainties $\sigma_{\hat \theta_k}$. 
The projection on the isochrones is performed as follows: 
\begin{enumerate}
\item
\label{step1_projection}
We select all of the isochrone points that have tabulated values within $n\cdot \sigma_{\hat \theta_k}$ from each considered observed parameter, where $n$ is arbitrarily chosen by the user. If less than 50 points are selected, the range is expanded to  $3n\cdot \sigma_{\hat \theta_k}$. This usually happens when the observed uncertainties are largely under-evaluated. \\

\item 
For each selected node on the isochrones, $i$, we assign a Gaussian weight $w_i$, which depends primarily on its distance from the measured observables. In practice: 
\begin{equation}
w_i=p_i \cdot  \beta_i \cdot \exp \left ( -\sum_k \frac{(\theta_{i,k}-\hat \theta_k)^2}{2\sigma_{\hat \theta_k}^2} \right ),
\label{eq:isochrone_weights}
\end{equation}
where $\theta_{i,k}$ corresponds to the tabulated parameters of the isochrones. 
The factor $\beta_i$, suggested by \citet{Zwitter10}, is defined as the stellar mass difference between two adjacent points on the same isochrone. It is equivalent  to assuming a flat prior on stellar mass, and it  provides higher probability to the likelihood of observing dwarfs,  reflecting their slower evolutionary phases (see panel 4 of Fig.~\ref{fig:isochrones}).  
The parameter $p_i$ is the prior on the age. It is equal to one if a flat prior is considered, or it can be a more complex function depending on the a priori knowledge of the investigated stellar population. The one adopted in this work is shown in Fig.~\ref{fig:age_prior}, and is discussed in Sect.~\ref{sec:age_prior}. \\

\item
The  projected parameters $\theta^*_k$, are  obtained by computing the weighted mean of the selected set of tabulated stars with parameters $\theta_{k}$: 

\begin{equation}
\theta^*_k=\frac{\sum_i w_i \cdot \theta_{k,i}}{\sum_i w_i}.
\end{equation}

\item

Finally, the associated uncertainty is obtained by computing the weighted dispersion : 
\begin{equation}
\sigma_{\theta^*_k}=\sqrt{ \frac{\sum_i w_i \cdot (\theta^*_k - \theta_{k,i})^2}{\sum_i w_i} }.
\label{eq:weighted_uncertainty}
\end{equation}
\end{enumerate}

Because in practice the method computes as many weights $w_i$ as points on the isochrones, restricting the computation to the points within $4\cdot \sigma_{\hat \theta_k}$  from the measurements (Step \ref{step1_projection}, above) ensures a significant speed-up of the algorithm without losing any contribution from isochrone points that are far from the input data. 
In addition, Step \ref{step1_projection} also ensures that no age, mass, etc is returned if the measured  parameters are too far from the set of isochrones. This is particularly crucial for the very low metallicity stars (\meta$<-2.2$\dex), since  a \teff-\logg\ match could, in principle, be found, despite being far from the edge of the grid of isochrones. 

Eventually, the algorithm returns as many parameters as the ones tabulated on the isochrones. We save, however, only the projected \teff, \logg~and \meta, the age ($\tau$), the initial stellar mass ($M_{\rm ini}$),  the absolute magnitudes $G$, \BP, \RP~and their associated uncertainties.  From the projected magnitudes, one can also infer the interstellar reddening,  $\EBPRP$, and the  interstellar extinction, $A_G$ (see Sect.~\ref{sec:extinctions}).

We note that the difference between the input \teff, \logg~and \meta\ and the output (i.e. projected) ones can serve as diagnostics of the projection. Typically, when the difference is too large\footnote{Empirically, when dealing with real Gaia data, we have found that thresholds of 200\,K in \teff, 0.3\dex\ in \logg, and 0.1\dex\ in \meta\ are enough to remove outliers. Other values, though, may be selected by the users depending on the application.}, this implies that no isochrones are found nearby the input values, therefore questioning the reliability of the output parameters (and perhaps the input ones as well).

\subsection{Choice of an age prior}
\label{sec:age_prior}

\begin{figure}[t!]
\begin{center}
\includegraphics[width=0.8\linewidth, angle=0]{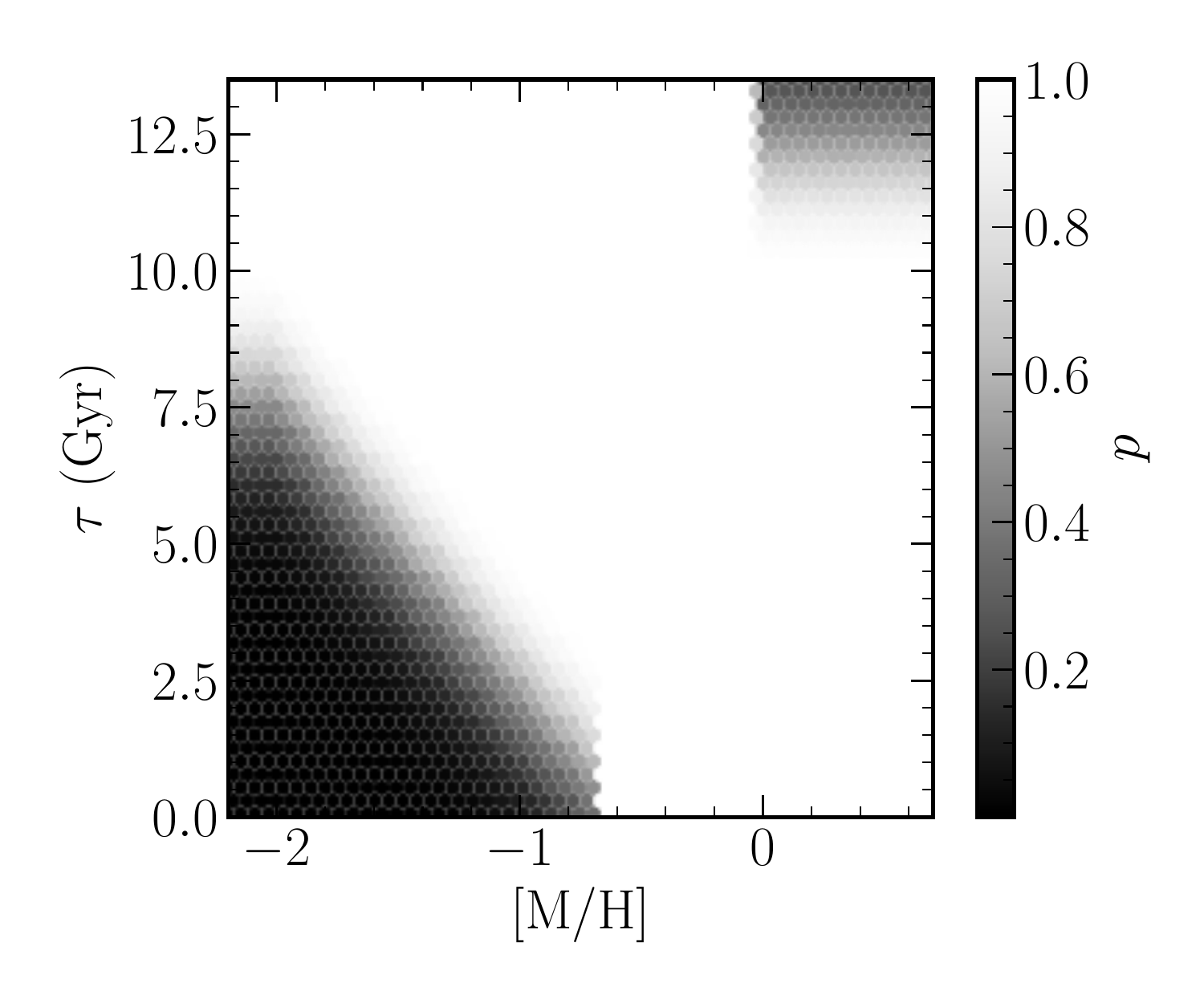}
\caption{ Age priors, $p$, adopted in this work, used in Eq.~\ref{eq:isochrone_weights}. These are half-Gaussians of $\sigma=2\Gyr$, centred on different ages depending on the metallicity.   }
\label{fig:age_prior}
\end{center}
\end{figure}

Unlike some isochrone projection methods available in the literature (some of them optimised to only get the line-of-sight distances, see Sect.~\ref{sec:Introduction}), the one we present in this paper does not adopt a specific age prior based on the galactic populations  (see Fig.~\ref{fig:age_prior}). 
As a matter of fact, for the majority of the stars ($-0.7\leq \meta \leq0$\dex) a flat prior is adopted. This choice is motivated by the facts that the definitions of thin and thick discs are not as clear as they used to be  \citep[chemical vs geometrical  discs, see for example][]{Hayden17}, that Gaia-Enceladus-Sausage spatial and chemical distribution is complex and partly entangled with the disc \citep[see for example][]{Belokurov18, Helmi18, Myeong19, Feuillet20}, and that the properties of many of the other accreted populations highlighted with Gaia show that a given locus in physical and chemical spaces  is a mix of several different populations of different ages \citep[e.g., ][and references therein]{Kordopatis20, Laporte20, Naidu20, GaiaAnticenter21}.

 Such a flat age prior, however, is poorly justified for super-solar metallicity or metal-poor  stars, for which one can safely suppose that the former cannot be too old \citep[metal-rich bulge stars are $\sim9 \Gyr$, e.g.][]{Schultheis17, Bovy19} and that the latter cannot be too young \citep[the metal-poorest thin disc stars in the Solar neighbourhood are found to be  $\sim8-9\Gyr$ old whereas the thick disc is found to be older than $7-8\Gyr$, see e.g.][]{Bensby14, Haywood18}.   
  For this reason we adopt half-Gaussian weights,  for $\tau>\tau_c$ if $\meta>0$ or $\tau<\tau_c$ if $\meta<-0.7\dex$:
   \begin{equation}
p=\exp\left(-\frac{(\tau-\tau_c)^2}{2\cdot \sigma_\tau^2}\right).
\end{equation}
We take $\tau_c=10\Gyr$ for $\meta>0\dex$ and $\meta<-2\dex$, i.e. a star has less probabilities of being older than $10\Gyr$ if it has a super-solar metallicity and less probabilities of being younger than $10\Gyr$ if it is more metal-poor than $-2\dex$.  
For stars with $-2<\meta<-0.7\dex$, we adopt: 
\begin{equation}
\tau_c=-5.4\cdot \meta -0.8,
\end{equation}
which imposes $\tau_c=3\Gyr$ at $\meta=-0.7\dex$ and $\tau_c=10\Gyr$ at $\meta=-2\dex$.
The resulting weights as a function of metallicity are shown in Fig.~\ref{fig:age_prior}.

%%%%%%%%%%%%%%%%%%%%%%%
\subsection{Validation of the method on synthetic data}
\label{sec:synthetic_tests}
In order to evaluate the method's performance, we first test it on a set of synthetic data. We randomly select, amongst the entire isochrone set (Sect.~\ref{sec:isochrones_set}), 
200 turn-off stars (defined as $5000 \leq$ \teff$\le8500$\,K and \logg$\ge 3.5$), 200 Red Giant Branch (RGB)  stars (defined as \teff$\le6000$\,K and \logg$< 3.5$) and 200 main-sequence stars (defined as \teff$<4800$\,K and \logg$\ge 4.0$). 
We note that this sample contains non-realistic stars, e.g. metal-rich stars older than 12\Gyr. We make sure, however, not to select stars younger than 5\Gyr~ with metallicities lower than $-1$\dex.

We consider four different types of projection, labelled as follows: 
\begin{itemize}
\item {\tt spec}: projects only $\hat \theta_k=$\{\teff, \logg, \meta\}.
\item {\tt speck}: projects \{\teff, \logg, \meta, $K_s$\}.
\item {\tt specjhk}: projects \{\teff, \logg, \meta, $J$, $H$, $K_s$\}.
\item {\tt specjhkg}: projects \{\teff, \logg, \meta\ $J$, $H$, $K_s$, $G$\}.
\end{itemize}

In addition, we consider different test-cases, where the value of each of the input parameters is randomised according to a normal distribution of standard deviation associated to a given uncertainty (see Table~\ref{Tab:simulation_input_uncertainties}). The sample labelled `Q25'  adopts the parameter uncertainties of the 25th percentile of the \gspspec~ catalogue for a given specific parameter.  Similarly, Q50 and Q95 adopt the uncertainties of the 50th and 95th percentiles. 
We note that  the uncertainties for the photometric filters are estimated as the quadratic sum of the uncertainty of the apparent magnitude in that filter and the uncertainty on the absolute magnitude derived by the distance modulus\footnote{We note, that main sequence stars with large uncertainties on their atmospheric parameters usually have small uncertainties in their distance modulus as they are  relatively nearby.} (see Sect.~\ref{sec:projection_flavours_description} for further details).

\begin{table}[t!]
\caption{Uncertainty percentiles for the \gspspec~sample adopted for the performance tests of the projection method.}
\begin{center}
\begin{tabular}{c|cccc}
parameter & Q25 & Q50 & Q75 & Q95 \\ \hline \hline
\teff\ (K)&35&  70& 130&  350\\
\logg\ ($g$ in \cgs)&0.1& 0.2&  0.35& 0.5\\
\meta\ (dex)&0.05& 0.1& 0.2& 0.4\\
$G$ (mag) &0.015& 0.03& 0.06& 0.14\\
$J$, $H$, $K_s$ (mag)&0.03& 0.04& 0.07& 0.15\\
 \hline
\end{tabular}
\end{center}
\label{Tab:simulation_input_uncertainties}
\end{table}%

\begin{figure*}[ht!]
\begin{center}
\includegraphics[width=\linewidth, angle=0]{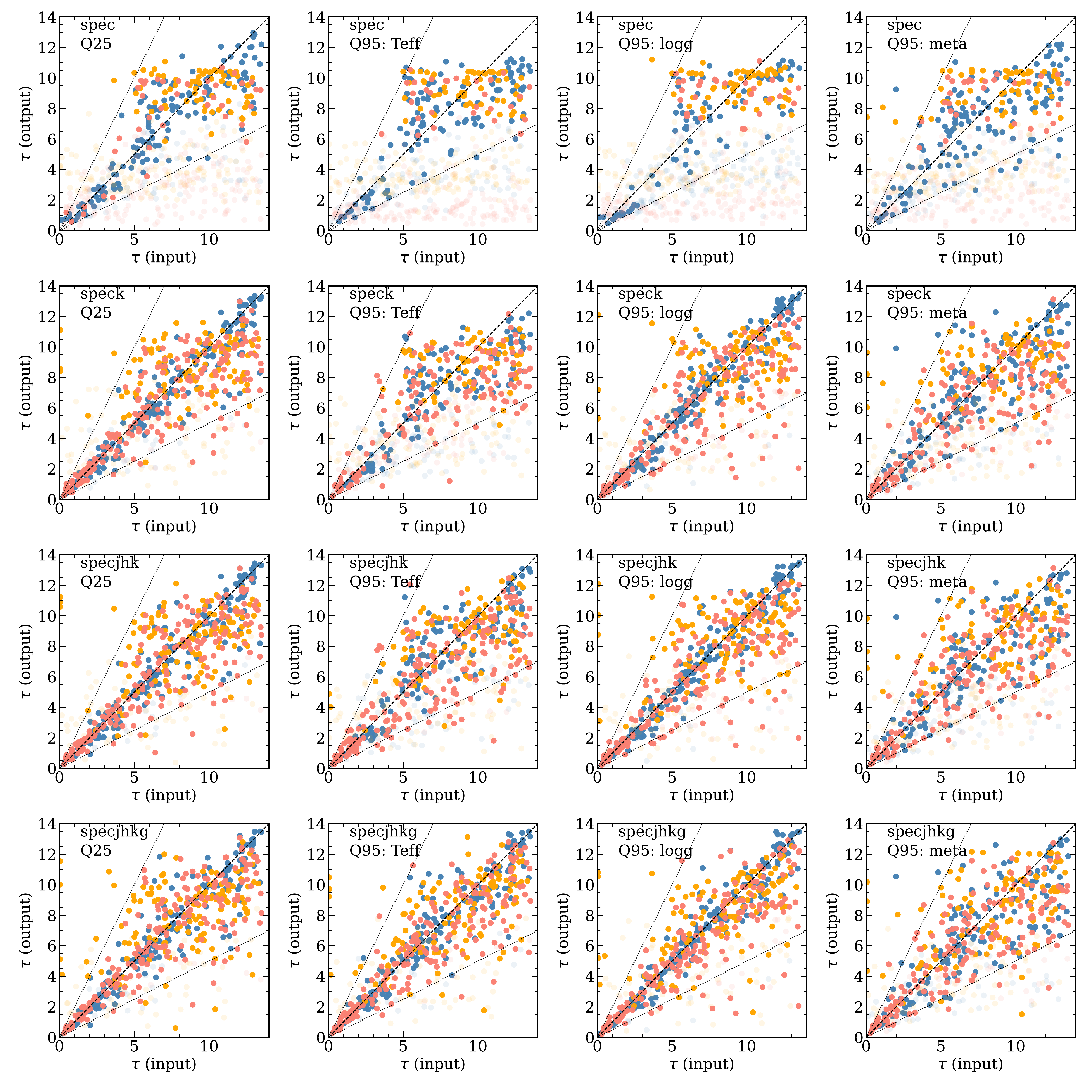}
\caption{ True versus output ages (in Gyr) for the test sample of synthetic data. 
Blue, orange and red points represent  turn-off, main-sequence and red giant stars, respectively. 
The dashed black line represents the 1:1 relation, on which a star with perfectly well estimated age should be on.  Dotted lines show deviations from perfect estimation of $\pm 50$ per cent.
Input parameters are randomised according to the Q25 uncertainties (see Table~\ref{Tab:simulation_input_uncertainties}), unless specified at the top left corner. 
 First row plots consider a projection of only \teff, \logg\ and \meta~ (labelled {\tt spec}), whereas rows two, up to four consider in addition the $K_s$ magnitude ({\tt speck}), the $J,H, K_s$ magnitudes ({\tt specjhk}), and the  $J,H, K_s, G$ magnitudes ({\tt specjhkg}), respectively. Finally, solid points are the stars for which the estimated relative age uncertainty is smaller than 50 per cent, whereas the semi-transparent ones have estimated relative age uncertainties larger than 50 per cent. See Sec.~\ref{sec:synthetic_tests} for further details. 
}
\label{fig:Synthetic_projection_ages}
\end{center}
\end{figure*}

\begin{figure*}[ht!]
\begin{center}
\includegraphics[width=\linewidth, angle=0]{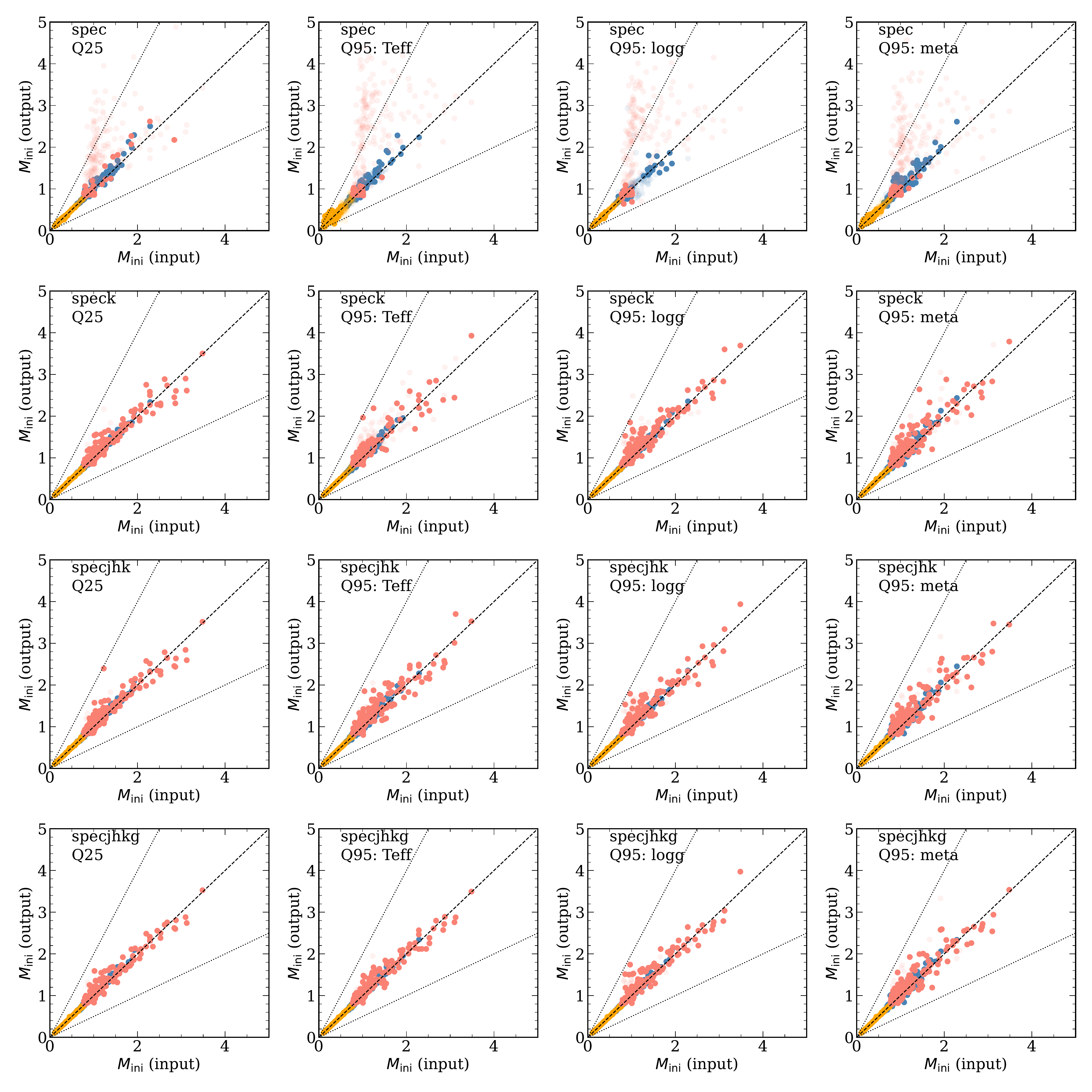}
\caption{ Same as Fig.~\ref{fig:Synthetic_projection_ages} for the initial stellar masses. Semi-transparent points are the ones having estimated relative age uncertainties greater than 50 per cent, whereas the filled circles have estimated age uncertainties smaller than 50 per cent. }
\label{fig:Synthetic_projection_masses}
\end{center}
\end{figure*}

Figures~\ref{fig:Synthetic_projection_ages} and \ref{fig:Synthetic_projection_masses}  compare the real and the derived ages and masses,  where each row shows a different projection flavour. Q25 uncertainties are adopted for all of the parameters, except for columns two to four, where one parameter each time adopts Q95, as indicated at the top left corner (while keeping the other parameters at Q25). Solid circles are the stars for which the estimated relative uncertainty in age is smaller than 50  per cent, whereas the semi-transparent circles have uncertainties larger than 50 per cent.

These figures show that, provided we apply a 50 per cent filter on the estimated  relative age uncertainty,  the ages for turn-off stars  are always well determined (blue points), even when input parameter uncertainties are large (Q95). On the other hand, giants and main-sequence stars (red and orange points, respectively)  require additional information to be used during  the projection (which can be in the form of absolute magnitudes or age priors)  in order to have their ages well derived. When this is done, {\tt speck}, {\tt specjhk} and {\tt specjhkg} return reliable estimations (age errors smaller than 50 per cent) for all of the stellar types, even with Q95 input uncertainties.

The mass estimations for main-sequence and turn-off stars are always reliable, regardless of the projection-flavour or input parameter uncertainties. This is not the case for giants, for which the {\tt spec} projection overestimates the values. Furthermore, we find that the mass uncertainties are generally underestimated, and that a filtering on the age uncertainty removes more efficiently the outliers, at the cost of removing simultaneously the mass estimations for the main-sequence stars. 
In what follows, we decide to filter-out the masses of solely the giants only if their relative age uncertainty is larger than 50 per cent.

Finally, we also investigated the effect of magnitude uncertainties, as well as interstellar reddening on the output parameters. For this test-case we set uncertainties for \teff, \logg~ and \meta~ always  equal to Q50, whereas we adopted uncertainties for $J$, $H$, $K_s$ and $G$ of Q25 and Q95. 
Three different  extinction values were considered:  $A_V=0.0$\,mag , $A_V=0.2$\,mag and $A_V=2.5$\,mag (converted for each filter, see Sect.~\ref{sec:ages_compilation}). 
Results for the ages are shown in the Appendix~\ref{appendix:projection_extinction}. In short, 
{\tt speck} projection always gives good ages and masses, even with Q95 and $A_V=2.5$\,mag. 
We find that $A_V=0.3$ introduces small biases for {\tt specjhk} and {\tt specjhkg} if the uncertainties on the magnitudes are small (i.e. Q25), but this bias mostly disappears for larger uncertainties (Q95).

%%%%%%%%%%%%%%%%%%%%%%%%%%%%%%%%
\begin{figure*}[t!]
\begin{center}
\includegraphics[width=\linewidth, angle=0]{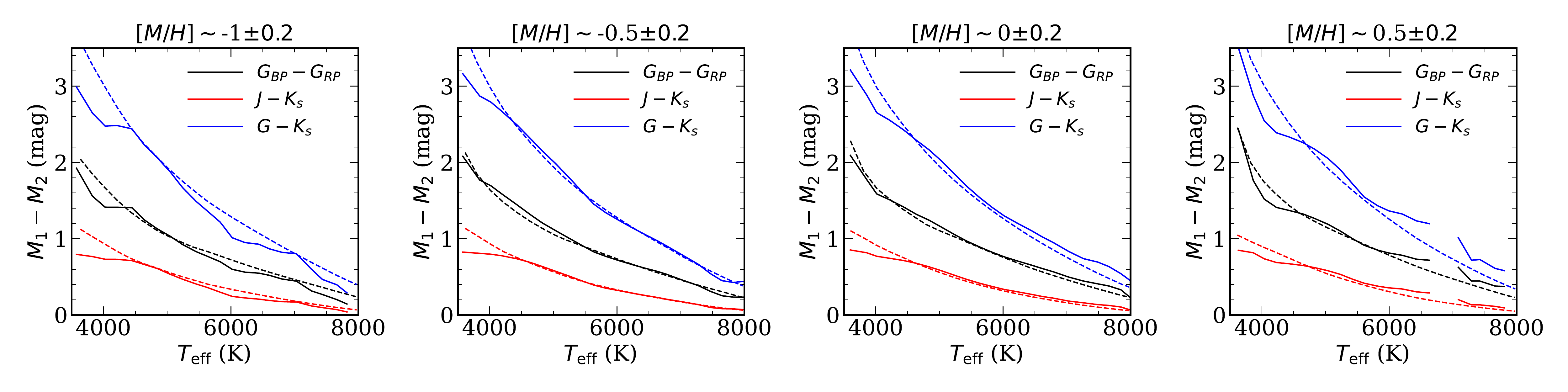}
\caption{Relation between \teff\  and \BP$-$\RP~(black), $J-K_s$ (red), $G-K_s$ (blue) for the adopted isochrones (in dashed lines), and the Gaia-RVS observed stars (solid lines) within 150\,pc from the Sun (to minimise the extinction effects) at four different metallicity regimes.   }
\label{fig:Teff_coulour_calibration}
\end{center}
\end{figure*}

\subsection{Specific considerations for projection of real data }
\label{sec:real_life}
When projecting real Gaia data on the isochrones, some specific points, that were not considered in the previous sections, need to be taken into account. These, concern the colour-calibration of the isochrones, and the $\alpha-$ enhancement of the stars. We discuss these points below.

\subsubsection{Colour-\teff~calibration}

For a reliable projection on the isochrones (or if one desires to use the output \BP~ and \RP~magnitudes, see Sect.~\ref{sec:extinctions}),  the relation between the \teff\ and the colour indexes must be the same for the observed values (in our case, \gspspec) and  for the ones on the isochrones. Following \citet{Santos-Peral21}, we investigated the relation between these two, for four different metallicity regimes ($-1, -0.5, 0, +0.5\dex$) for stars closer than 150\,pc from the Sun, in order to minimise the effect of reddening on the observed colours. Results shown in Fig.~\ref{fig:Teff_coulour_calibration} suggest no significant differences  between the isochrone and the observed relations, at least for stars hotter than 4000\,K, therefore implying that no specific calibration needs to be performed on the isochrone colours in order to be on the same scale as  \gspspec's \teff.

\subsubsection{Effect of non-solar $\alpha$-enhancement }
\label{sec:Salaris}
PARSEC isochrones do not include any variation of $\alpha-$abundance. Indeed, all of the sets adopt solar value. In order to evaluate the effect of non-solar  $\alpha$-enhancement, we used  the \citet{Salaris93} formula, as derived in \citet{Valentini19},  to  convert \gspspec's calibrated \meta~ to a proxy for the overall metallicity including $\alpha-$elements, $\meta_\alpha$.
The adopted formula, is the following: 
\begin{equation}
\label{eq:Salaris}
\meta_\alpha=\meta_{\gspspec}+\log_{10}(C\cdot 10^{\afe}+(1-C))
\end{equation} 
with C$=0.661$, and with $\afe$\ the calibrated value provided by \gspspec~obtained with a third degree polynomial (see first row of Table~4 of \citealt{GaiaDR3-GSPspec}).

Using the {\tt spec} projection, adopting one or the other metallicity estimate, resulted to differences  smaller than $1\Gyr$ and  $0.1M_\odot$ for 90 per cent of the stars, for $\tau$ and $\Mini$, respectively. 

Given that these differences are smaller than the uncertainty on the derived ages and masses, and to the fact that $\afe$ abundances have a precision of the order of $\sim0.07$\dex~(that can become larger depending on the quality of the spectra, see \citealt{GaiaDR3-GSPspec}), we decided in what follows not to use $\meta_\alpha$ . However, projections obtained using  $\meta_\alpha$ can be provided, upon request to the first author.

%%%%%%%%%%%%%%%%%%%%%%%%%%
\subsection{Gaia DR3 application: Computation of the datasets}
\label{sec:dataset_definitions}

\subsubsection{Projection flavours, input parameters and associated uncertainties}
\label{sec:projection_flavours_description}
We performed four different isochrone projections (leading therefore to four sets of ages, masses, interstellar extinctions), according to the flavours described in Sect.~\ref{sec:synthetic_tests}: {\tt spec}, {\tt speck}, {\tt specjhk} and {\tt specjhkg}. 

The atmospheric parameters are the calibrated \gspspec~ ones, following Eq.~1 (for \logg) and Eq.~2 (for \meta, with a fourth degree polynomial) of \citet{GaiaDR3-GSPspec}.  
The absolute magnitudes in the $G$, $J$, $H$ and $K_s$ bands have been obtained using the distance modulus, adopting the \citet{Bailer-Jones21} geometric distances ($r$) that assume a distance prior along with the parallax information, and neglecting entirely the line-of-sight extinction. Although this assumption has no consequence for the {\tt spec} projection, neglecting  the line-of-sight extinction introduces an increasing amount of bias in the output ages and masses the bluer the central wavelength of the filter (i.e. in increasing order: $K_s$, $H$, $J$ and lastly $G$, see Appendix~\ref{appendix:projection_extinction}).  

The uncertainties required in Eq.~\ref{eq:isochrone_weights} for the projection, are obtained  assuming Gaussian  posteriors for \gspspec~\teff, \logg~and \meta~(i.e. taking half of the difference between the upper and lower confidence value). 
The uncertainty on the absolute magnitudes $M_\lambda$, for a given filter $\lambda$, are obtained as the quadratic sum of the uncertainty on the observed magnitude, $m_\lambda$, and the uncertainty term coming from the distance modulus $\mu$: 
\begin{equation}
\sigma_{M_\lambda}=\sqrt{ \sigma^2_\mu + \sigma^2_{m_\lambda} },
\label{eq:abs_mag_uncertainty}
\end{equation}
where $\sigma_\mu$ is the uncertainty on the distance modulus derived using standard error analysis: 
\begin{equation}
\sigma_\mu=\frac{\sigma_r}{0.461\cdot r},
\end{equation}
with $\sigma_r$   the (assumed) Gaussian uncertainty on the distance ($r$), obtained also as the half of the difference between the upper and the lower confidence value of $r$.

\begin{figure}[ht!]
\begin{center}
\includegraphics[width=\linewidth, angle=0]{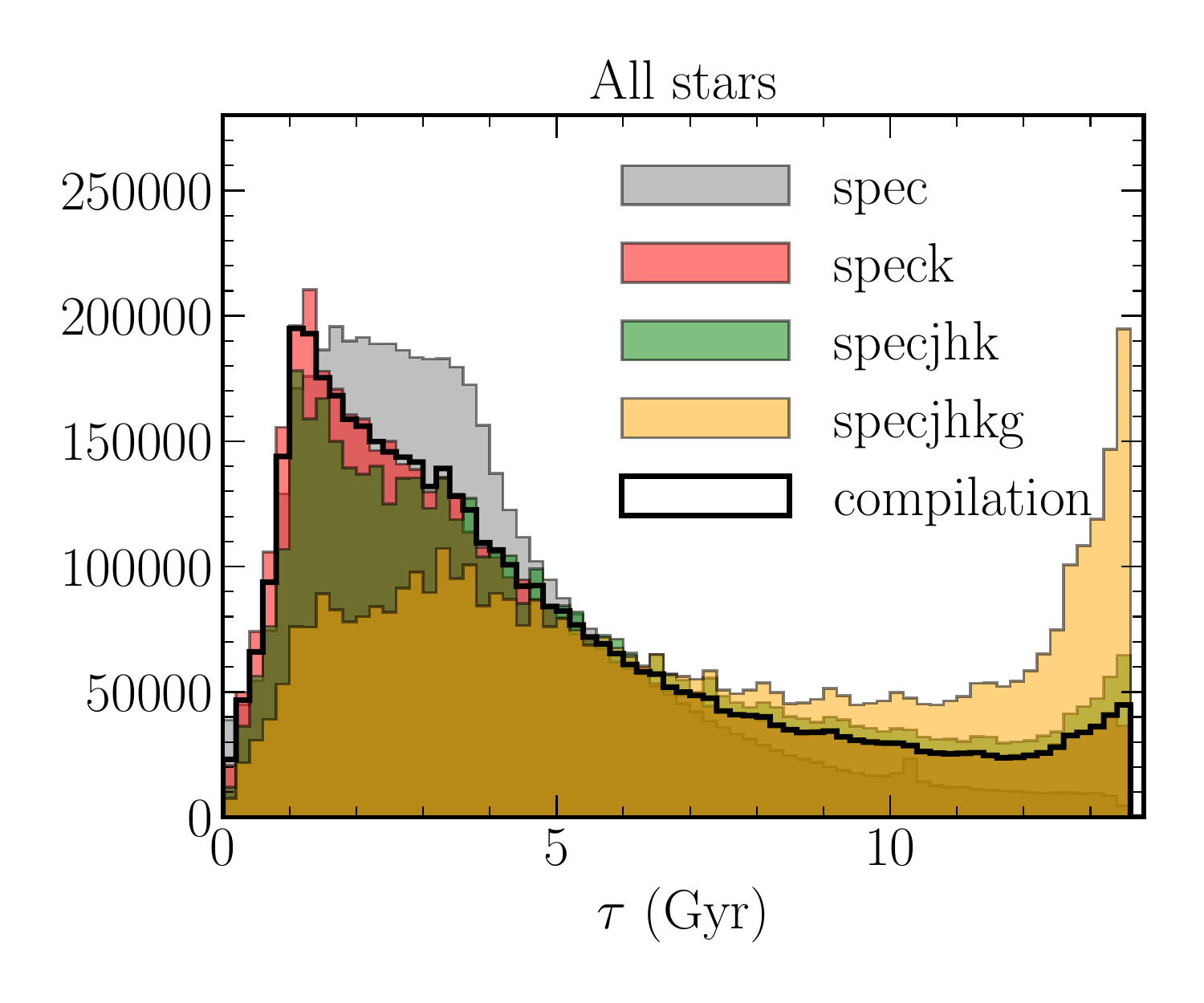}\\
\includegraphics[width=0.95\linewidth, angle=0]{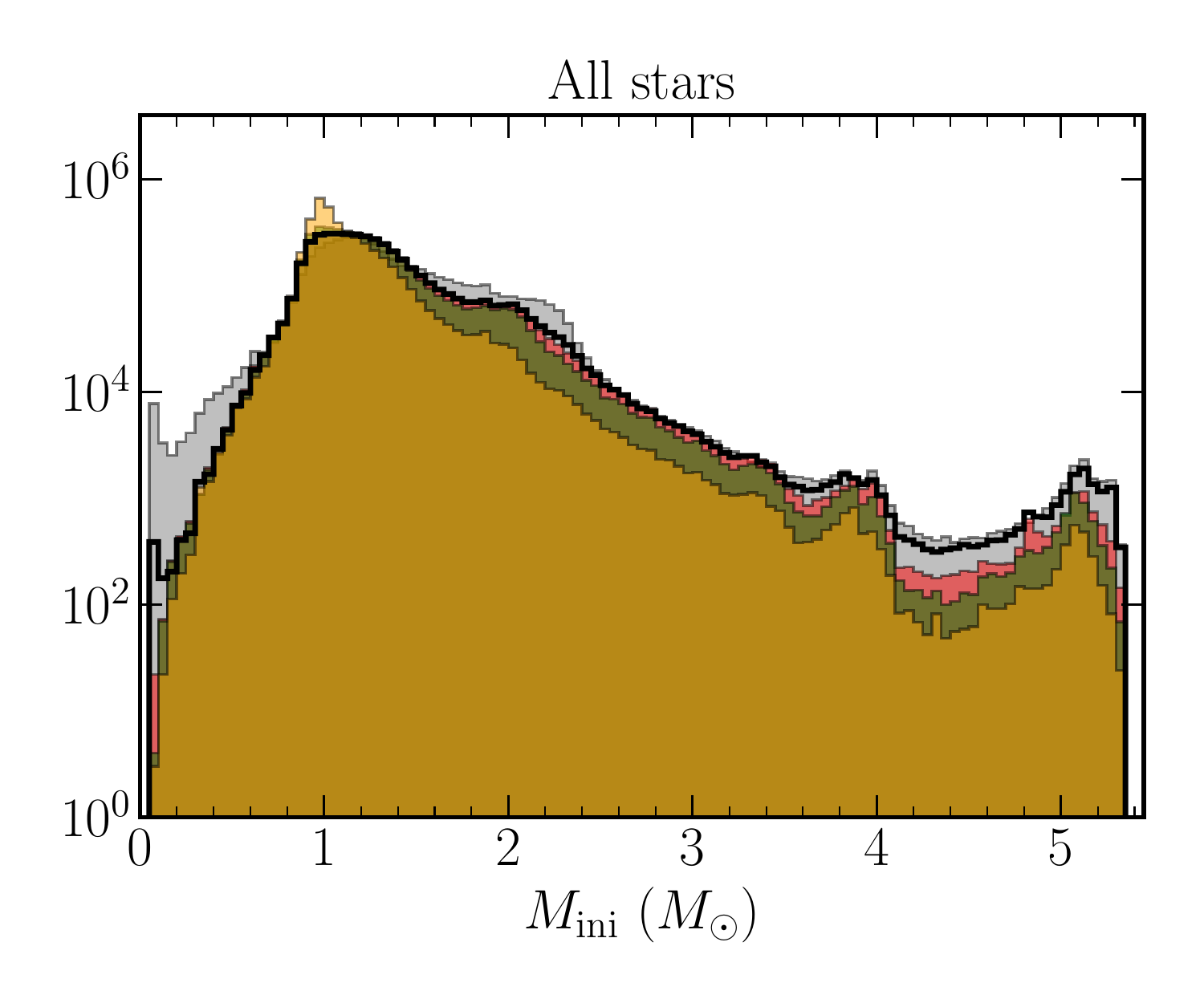}
\caption{Histograms of age (top) and initial mass (bottom, in logarithmic scale) of each projection flavour and for the compiled sample, for all the \gspspec~stars that have a {\tt KMgiantPar\_flag $= 0$}.
}
\label{fig:Ages_flavours}
\end{center}
\end{figure}

Figure~\ref{fig:Ages_flavours} compares the age (top) and mass (bottom) distributions of the different projections. Each distribution contains all of \gspspec's stars, except those  having  {\tt KMgiantPar\_flag $\ne 0$}, for which \teff\ and \logg\ are less reliable  \citep[see][]{GaiaDR3-GSPspec}. 
 One can see that the {\tt specjhkg} projection finds significantly older stars, especially compared to the {\tt spec} projection, the latter showing the youngest ages, on average, with a broad peak at $\tau\sim3.5\Gyr$. 
 These differences can be understood given the analyses in the previous sections; 
  on the one hand, as extinction is not taken into account in the {\tt speck}, {\tt specjhk} and {\tt specjhkg} projections, giants with reddened colours tend to lie on older isochrones, biasing the age estimation towards larger values. As shown in Appendix~\ref{appendix:projection_extinction}, this bias is stronger the bluer the  filter, eventually leading to the peak at $\tau\sim12\Gyr$ for the {\tt specjhkg} projection. On the other hand, the spectroscopic projection, as shown in Fig.~\ref{fig:Synthetic_projection_ages}, tends to be biased towards younger ages for all stellar types except the turn-off stars.

 \subsubsection{Catalogue compilation}
 \label{sec:ages_compilation}
 The differences seen in Fig.~\ref{fig:Ages_flavours}, naturally lead to the necessity of finding an optimal combination of projection flavour in order to get the most reliable ages and masses. 
 
 This was attempted by obtaining  the dust-reddening $E(B-V)$ maps from  \citet{Schlegel98}\footnote{This choice is motivated by the fact that other dust maps, e.g. \citet{Green19}, either do not cover the entire sky, or do not probe the entire  Gaia-RVS volume. }, using the Python script {\tt dustmaps}\footnote{\url{https://dustmaps.readthedocs.io/en/latest/}} \citep{Green18}, and correcting the most reddened regions ($E(B-V)>0.1$) as in \citet{Bonifacio00}. 
 These were then converted into $A_V$, assuming $R_V=3.1$, and then to $A_J$, $A_H$, $A_{Ks}$ and $A_G$, following the adopted PARSEC 3.5 extinction coefficients of  \citet{Cardelli89}, \citet{O'donnell94} and the Gaia EDR3 passbands of \citet{Riello21}, as summarised in Table~\ref{Tab:AV_convertion}.
 Finally, these extinctions were then  compared with the uncertainties of the absolute magnitudes in each respective  band, as derived from Eq.~\ref{eq:abs_mag_uncertainty}.

  \begin{figure*}[ht]
\begin{center}
\includegraphics[width=\linewidth, angle=0]{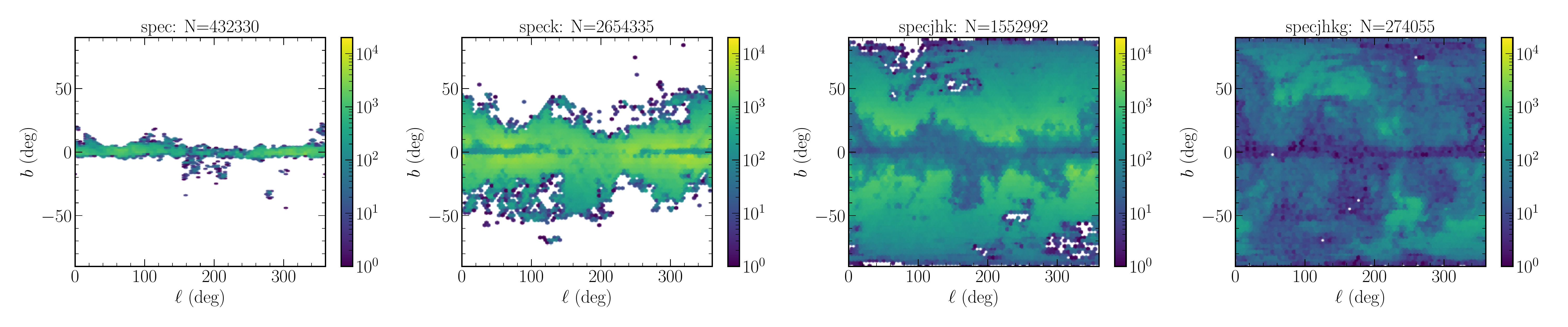}
\caption{Spatial distribution, in galactic sky-coordinates $(\ell,b)$, of the number of stars for which we adopt each projection flavour. The total of selected stars and the considered projection are indicated at the top of each plot. Only targets with \gspspec~{\tt KMgiantPar\_flag$=0$} are considered here. }
\label{fig:Ages_flavours_lb}
\end{center}
\end{figure*}

 \begin{table}[tp]
\caption{Extinction conversions, as adopted from PARSEC 3.5.}
\begin{center}
\begin{tabular}{c|cccc}
 & $A_J$ & $A_H$ & $A_{Ks}$  & $A_G$ \\ \hline
$A_\lambda/A_V$  & 0.28665 & 0.18082 & 0.11675 & 0.83278 \\
\end{tabular}
\end{center}
\label{Tab:AV_convertion}
\end{table}%

To compile our final, adopted, catalogue (labelled without any underscore in Table~\ref{tab:catalogue}), we selected, for a given star, the projection flavour according to the scheme below: 
%  For a given projection, if the uncertainty in the bluest magnitude  was smaller than twice the extinction in that band, then the parameters coming from that projection were adopted.
% In practice, we selected the results coming from:
 \begin{itemize}
 \item
 {\tt speck} where $A_{Ks}\le \sigma_{M_{Ks}}$, or $r\le300$\,pc and $A_V\le 3$\,mag.
 \item
  {\tt specjhk} where $A_{J}\le \sigma_{M_J}$, or $r\le100$\,pc.
  \item
  {\tt specjhkg} where $A_{G}\le \sigma_{M_G}$, or $r\le50$\,pc.
  \item
   {\tt spec} otherwise. 
  \end{itemize}
 
The resulting age and mass distributions of the compiled sample are shown in black in Fig.~\ref{fig:Ages_flavours}. 
 The conditions regarding the distance and $A_V$ are empirical, and take into account the fact that Schlegel's values are the total extinctions  along the line-of-sight, and that the Sun resides within a local bubble with very little extinction \citep{Fitzgerald68, Lallement14}.  The  extinction criterion is an educated guess resulting from the analysis of Sect.~\ref{sec:synthetic_tests} using reddened projections of synthetic data. Figure~\ref{fig:Ages_flavours_lb} shows the number of stars in bins of galactic $(\ell,b)$ where each projection flavour is adopted.  
 The {\tt spec} flavour is, as expected, selected only for stars within the galactic plane, and the {\tt speck} flavour is the one that contains the most of the stars.  Finally, the {\tt specjhkg} flavour contains the smaller number of stars, spread, however, all over $(\ell,b)$. 
 
 %% Add Kiel figures
 \begin{figure*}[ht!]
\begin{center}
\includegraphics[width=\linewidth, angle=0]{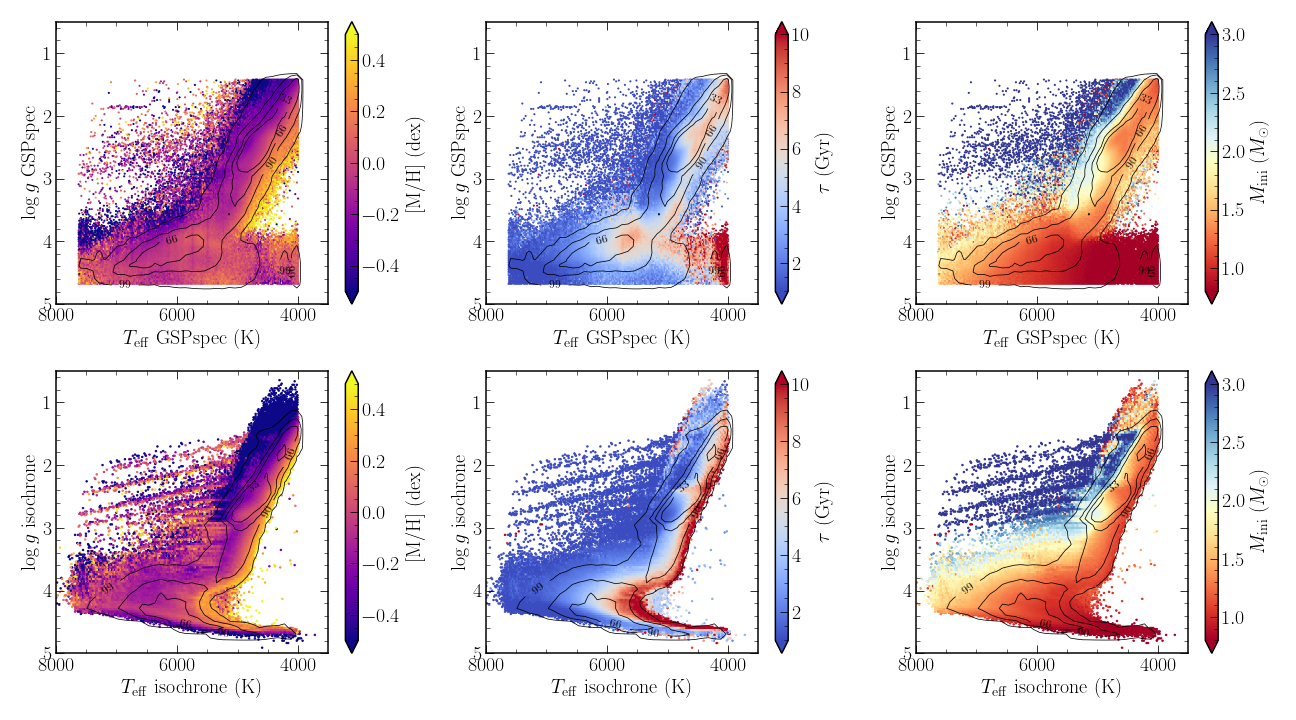}
\caption{ \teff-\logg~diagrams adopting \gspspec~parameters (top) and isochrone-projected parameters (bottom). Colour-code represents \gspspec~calibrated metallicity, age (in Gyr) and initial stellar mass (in $M_\odot$). Only targets with first 12 quality \gspspec~flags $\leq1$ and {\tt KMgiantPar\_flag$=0$} are considered here. }
\label{fig:Kiel_age_masses}
\end{center}
\end{figure*}

 Figure~\ref{fig:Kiel_age_masses} shows Kiel diagrams colour-coded by either metallicity, age or initial mass, with the input \teff~and \logg~(top row), and the output ones (bottom row).  These plots show that the parameters we recover in different regions of the diagram are compatible with what one would expect: low mass stars are mainly cool dwarfs, and old stars are at the redder part of the RGB. Similarly, the turn-off region contains significantly younger stars.

To conclude, we note that in the published catalogue (see Table~\ref{tab:catalogue}), the parameters resulting from this compilation are labelled without underscores. For the users that wish to perform their own combination of flavours, based on different criteria than the ones adopted here, the results from each separate flavour  are also published, with the appropriate and explicit label-name.

 \subsection{Validation of the ages and masses}
\label{sec:age_mass_validation}

\begin{figure}[ht!]
\begin{center}
\includegraphics[width=\linewidth, angle=0]{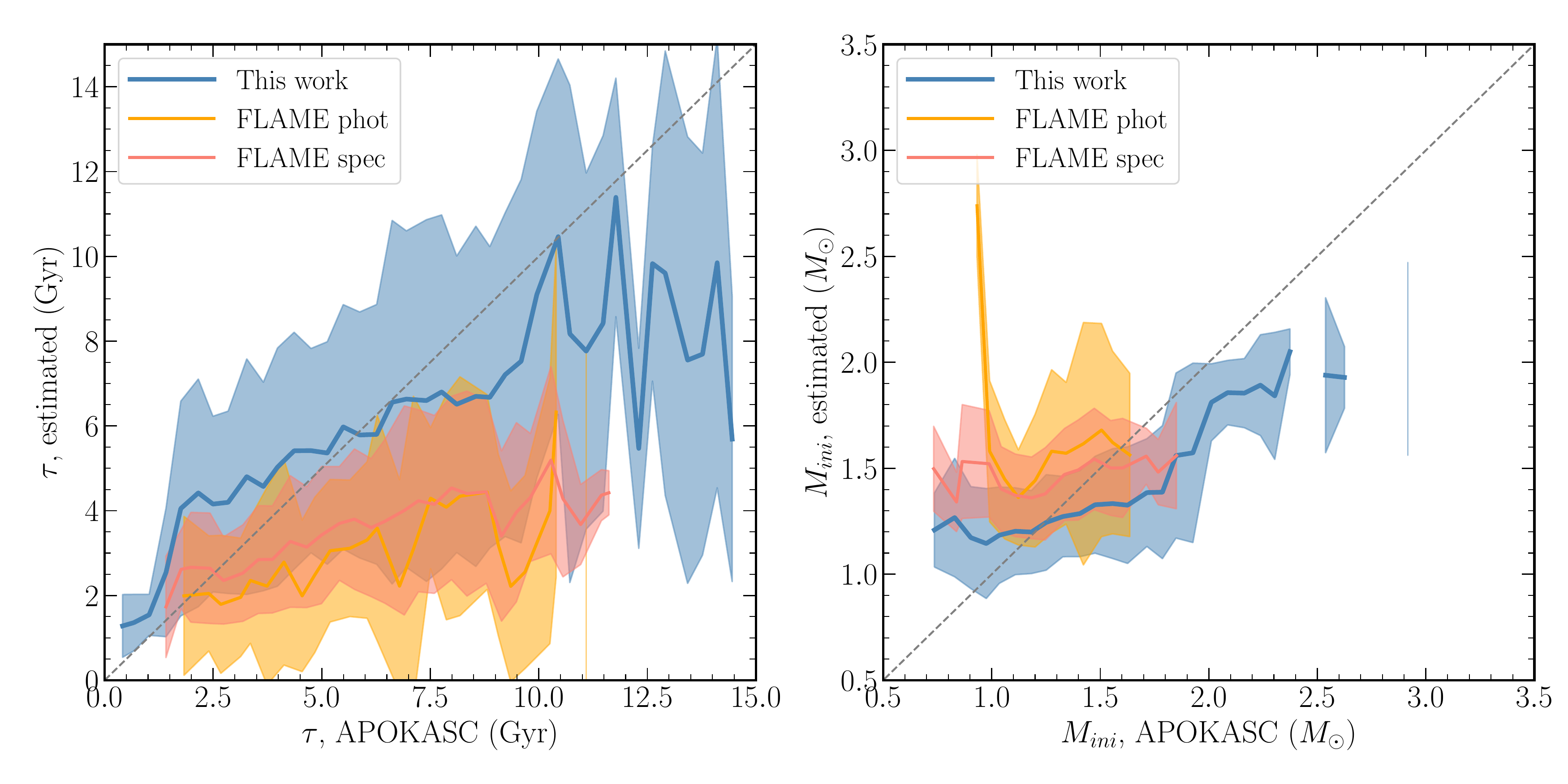}\\
\includegraphics[width=\linewidth, angle=0]{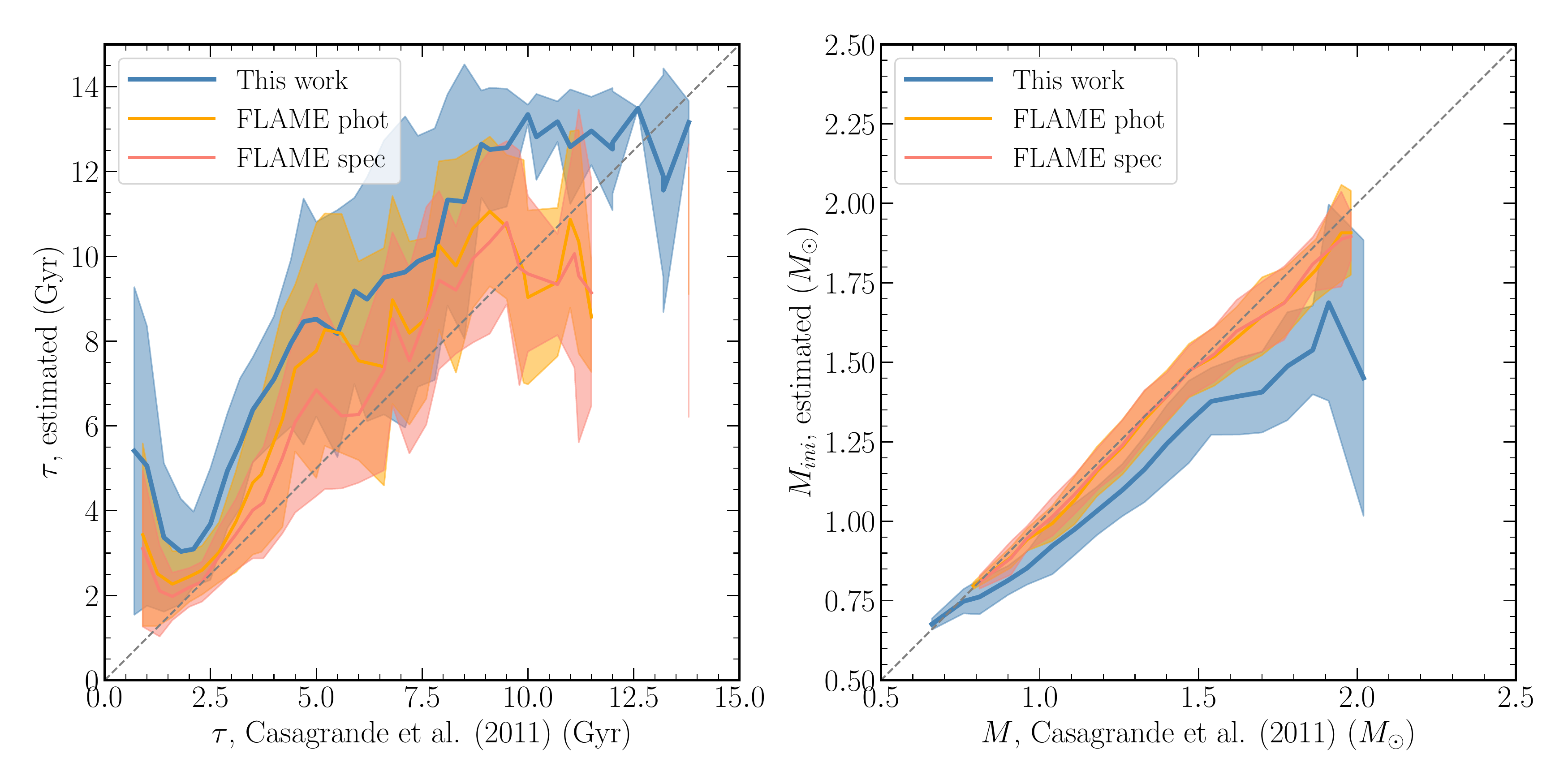}
\caption{Comparison of the derived ages (left) and masses (right) with the APOKASC-2 (top) and \citet{Casagrande11} (bottom) catalogues. Plots show running medians in bins of reference (catalogue) values.  In blue are shown the values derived with the compiled catalogue of Sect.~\ref{sec:ages_compilation}. In red and orange are shown the results of the FLAME-Spec and FLAME-Phot modules, respectively. Shaded areas represent the $\pm1\sigma$ around the running medians. Only the targets that have \gspspec~flags equal to zero and relative age uncertainties (on each module separately) smaller than 50 per cent are selected.  Additional cuts, imposing $\meta>-0.5\dex$, are applied for the FLAME results. }
\label{fig:APOKASC_comparisons}
\end{center}
\end{figure}

To validate our ages and masses we compare  our results with the APOKASC-2 asteroseismic values \citep{Pinsonneault18} that  include a wide metallicity range ($-2.5 \leq \meta \leq 0.5$\dex) but concern only giants (\logg$<3.3$), and the values from \citet{Casagrande11} that concern mostly main-sequence stars. We also compare the mean ages we derive for open clusters with the ones of \citet{Cantat-Gaudin20} and perform a similar exercise for globular cluster based on the selection of \citet{Gaia_Helmi18}.  Finally, we compare the values we obtain with the ones of the FLAME pipeline for the entire sample.  
In the subsections that follow, we always remove the stars that have a Renormalised Unit Weight Error ({\tt RUWE}) greater than 1.4, as they could potentially be non-single sources \citep[see][]{Lindegren_RUWE}, as well as the stars with \gspspec~{\tt KMgiantPar\_flag$>0$}. 

%%%%%%%%%%%%%%%%%%%%
\subsubsection{Comparison with APOKASC-2 and Casagrande et al. (2011) ages and masses}
\label{sec:APOKASC-Casagrande}
We required the 12 first \gspspec~ parameter quality flags to be smaller or equal to one (i.e. very good quality sample with the small atmospheric parameter uncertainties) and selected stars with  relative output age uncertainties  smaller than 50 per cent.  
The mean offset and standard deviations between the \gspspec~ atmospheric parameters and the literature ones are, for \teff, \logg~and \meta: $-20\pm100\,K$,  $0.01\pm0.21\dex$,  $-0.11\pm0.09\dex$  for the APOKASC-2 sample, and $-74\pm132\,K$,  $-0.06\pm0.17\dex$,  $-0.17\pm0.13\dex$ for the \citet{Casagrande11} sample. 

The blue lines in Fig.~\ref{fig:APOKASC_comparisons} show the running median in bins of reference catalogue values (i.e. APOKASC-2 or \citealt{Casagrande11}), for the compiled sample as defined in Sect.~\ref{sec:ages_compilation}. For comparison purposes, running medians obtained when considering the FLAME-Spec and FLAME-Phot datasets are also shown in red and orange\footnote{For the FLAME-Spec and FLAME-Phot sample, we also impose metallicities to be greater than $-0.5$\dex, as suggested by \citet{GaiaDR3-CU8-2}.}.  

One can see that we manage to reproduce adequately the literature ages, with mean offsets and standard deviations between the literature and the derived values of $(-0.77\pm3.4)\Gyr$ for APOKASC-2 and $(-2.4\pm2.6)\Gyr$ for \citet{Casagrande11}. These values drop to $(0.2\pm2.3)\Gyr$ and $(-1.5\pm2.3)\Gyr$ when we impose differences between the \gspspec~atmospheric parameters and the literature ones smaller than  $150\,K, 0.15\dex$ and $0.1\dex$ for \teff, \logg~and \meta, respectively.

We find that the ages for the giants tend to be slightly over-estimated until $\tau\sim11\Gyr$, above which value the median trend shows an underestimation of the ages. The first reason for this is that ages for old giants are inherently difficult to determine accurately because the isochrones are close to each other. Furthermore, since we do not have isochrones older than 13.7\Gyr, our code tends to sample  asymmetrically the ages, leading to a larger selection of  young isochrones compared to old ones. 
Additionally, we note that we have a relatively large disagreement on the ages of the youngest stars of \citet{Casagrande11}. Whereas it is not clear what is the origin of this large disparity, one can see that we also find a similar trend with FLAME-Spec and FLAME-Phot parameters, the later being based on completely different input parameters.

As far as the masses are concerned, we find a very good agreement, both for APOKASC-2 and \citet{Casagrande11}, with a slight overestimation of the masses for sub-solar mass giants, and a slight under-estimation for stars with masses greater than $\sim1.5\,M_\odot$. 
The mean offsets and standard deviations between the literature and the derived values are of $(0.00\pm0.3)\,M_\odot$ for APOKASC-2 and $(0.1\pm0.1)\,M_\odot$ for \citet{Casagrande11}. These values become $(-0.1\pm0.2)\,M_\odot$ and $(0.09\pm0.08)\,M_\odot$ when we impose differences between the \gspspec~atmospheric parameters and the literature ones smaller than  $150\,K, 0.15\dex$ and $0.1\dex$ for \teff, \logg~and \meta, respectively.

%%%%%%%%%%%%%%%%%%%%%%%%%%%%%
\subsubsection{Open cluster ages}
\begin{figure}[t!]
\begin{center}
\includegraphics[width=\linewidth, angle=0]{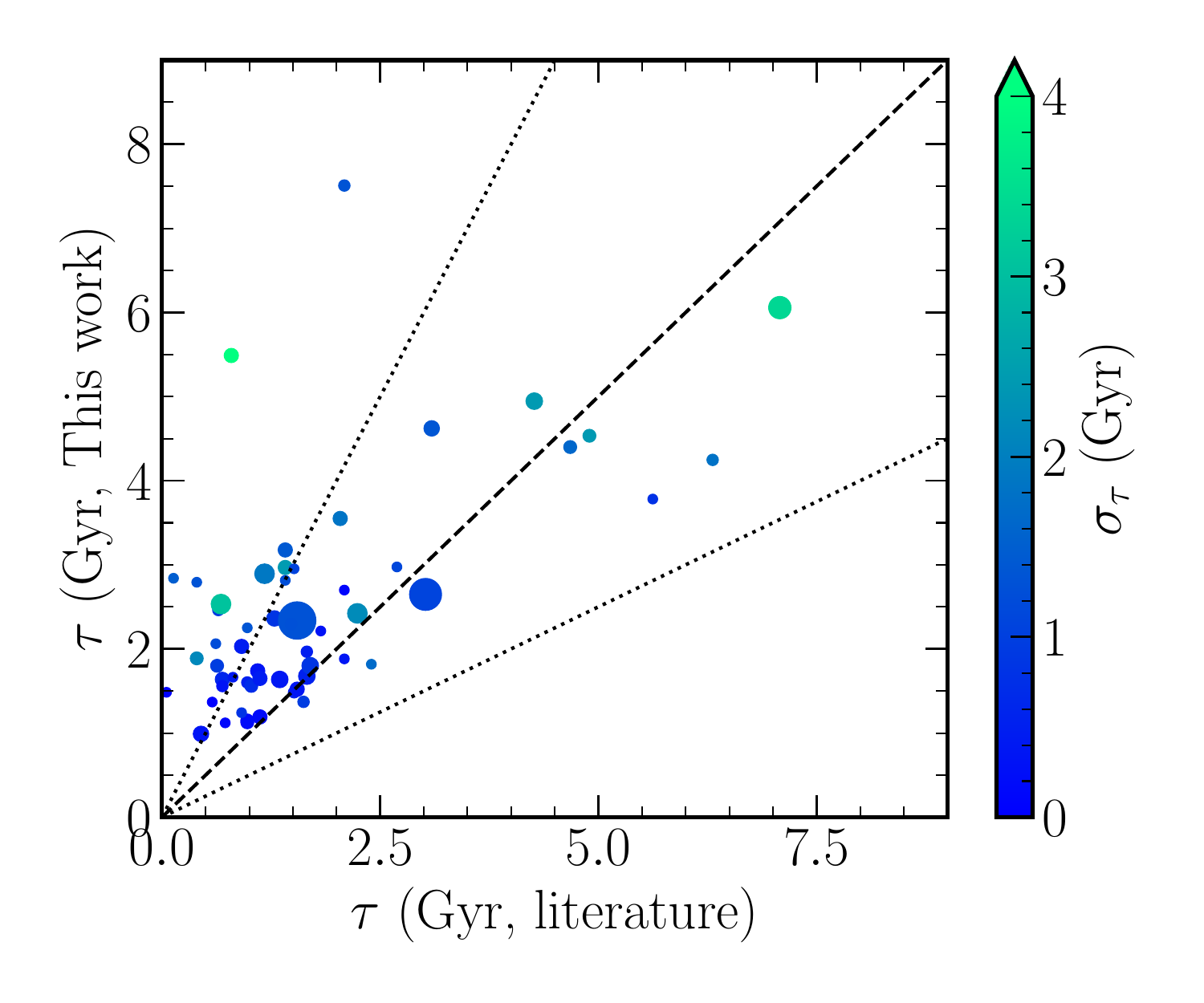}
\caption{Comparison of median open cluster ages with the ones of \citet{Cantat-Gaudin20}. 
The colour-code corresponds to the age dispersion that we obtain within the cluster, and the size of the points is proportional to the number of cluster members that made our selection. A minimum of three stars per cluster has been imposed. Dashed diagonal line represents the 1:1 relation, whereas the dotted lines show deviations from identity of $\pm 50$ per cent. }
\label{fig:cluster_comparisons}
\end{center}
\end{figure}

We selected from the sample of \citet{Cantat-Gaudin20} the stars that have a published probability higher than 95 per cent of being  part of a considered open cluster. We furthermore filtered out the stars that have \gspspec~flags greater than 1 and output relative age uncertainties greater than 50 per cent. From the resulting sample we computed the mean ages and dispersion per cluster and compared these numbers to the values published by \citet{Cantat-Gaudin20}. Results are shown in Fig.~\ref{fig:cluster_comparisons}. They suggest we perform very well for the young targets, with a small overestimation of the ages by $0.6\Gyr$ and an inner cluster age dispersion of the order of $\sim1\Gyr$ for most of the objects that contain at least some turn-off or subgiant stars. 

Amongst the few clusters for which the disagreement with the literature values is greater than 50 per cent (dashed lines in Fig.~\ref{fig:cluster_comparisons}), we can identify two different cases. On the one hand, we have the clusters that  contain mostly main-sequence stars, known to constrain poorly the age (see the case of Melotte25 in Fig.~\ref{fig:cluster_comparisons_isochrones}). 
On the other hand, we have the objects for which we suspect that the input parameters could be slightly biased, since our  proposed solution fits better the observed parameters than the isochrone with the literature age. 
These biases can either be due to a metallicity calibration that is different than for field stars \citep[see Fig.\,13 of][]{GaiaDR3-GSPspec} or due to large uncertainties on the distance modulus. 
The open cluster NGC2112 likely falls into this category, with a discrepancy with the literature estimations of  $\sim5\Gyr$  \citep{Carraro02, Kharchenko13, Cantat-Gaudin20}.

%%%%%%%%%%%%%%%%%%%%%%%%%%%
\subsubsection{Globular cluster ages}
\label{sec:globular_clusters}
We investigated the performance of our code for old and metal-poor stars, i.e. stars belonging to globular clusters. This regime is known to be difficult to get ages for field stars, as the isochrones, especially for giants are very close one to another. 
We selected the globular cluster targets compiled in \citet{Gaia_Helmi18} and cross-matched them with our sample, requiring \gspspec~quality flags on the atmospheric parameters to be less than or equal to 1. Furthermore, the median S/N of the selected spectra being low, of the order of $30$, we required that the differences between the input and output \teff, \logg~and \meta~to be less than  200\,K, 0.3~and 0.1\dex, in order to ensure that we have measurements close to the isochrones.  
For the six globular clusters that contained at least three stars fulfilling the conditions above, we computed the mean age, with the expectation of finding them older than $10\Gyr$ \citep[e.g.][]{VandenBerg13}. 

Despite the adopted age-prior for the metal-poor stars, we find ages ranging between $3.5\Gyr$ and $9.3\Gyr$, consistent with the results of Sect.~\ref{sec:APOKASC-Casagrande} for APOKASC-2, that suggest that old targets tend to have under-estimated ages. 
A closer investigation of  the Kiel and \teff-$M_{K_s}$ diagrams (see Fig.~\ref{fig:globular_cluster_comparisons_isochrones}) highlights the difficulty of  determining these ages properly due to the proximity of the various isochrones. 
On top of this challenge, we find that the mean metallicity of the stars within each cluster tend to differ by $\sim+0.15\dex$ compared to, for example, the values reported by \citet{VandenBerg13}. This metallicity difference, also noted in \citet{GaiaDR3_PVP_chemicalCartography}, without, however suggesting a specific calibration for these objects, is also partially responsible for the age offsets we find, compared to expectations.

%%%%%%%%%%%%%%%%%%%%%%%%%%%%%%%%%%%%%
\subsubsection{Comparison with FLAME ages and masses}
\begin{figure}[t!]
\begin{center}
\includegraphics[width=\linewidth, angle=0]{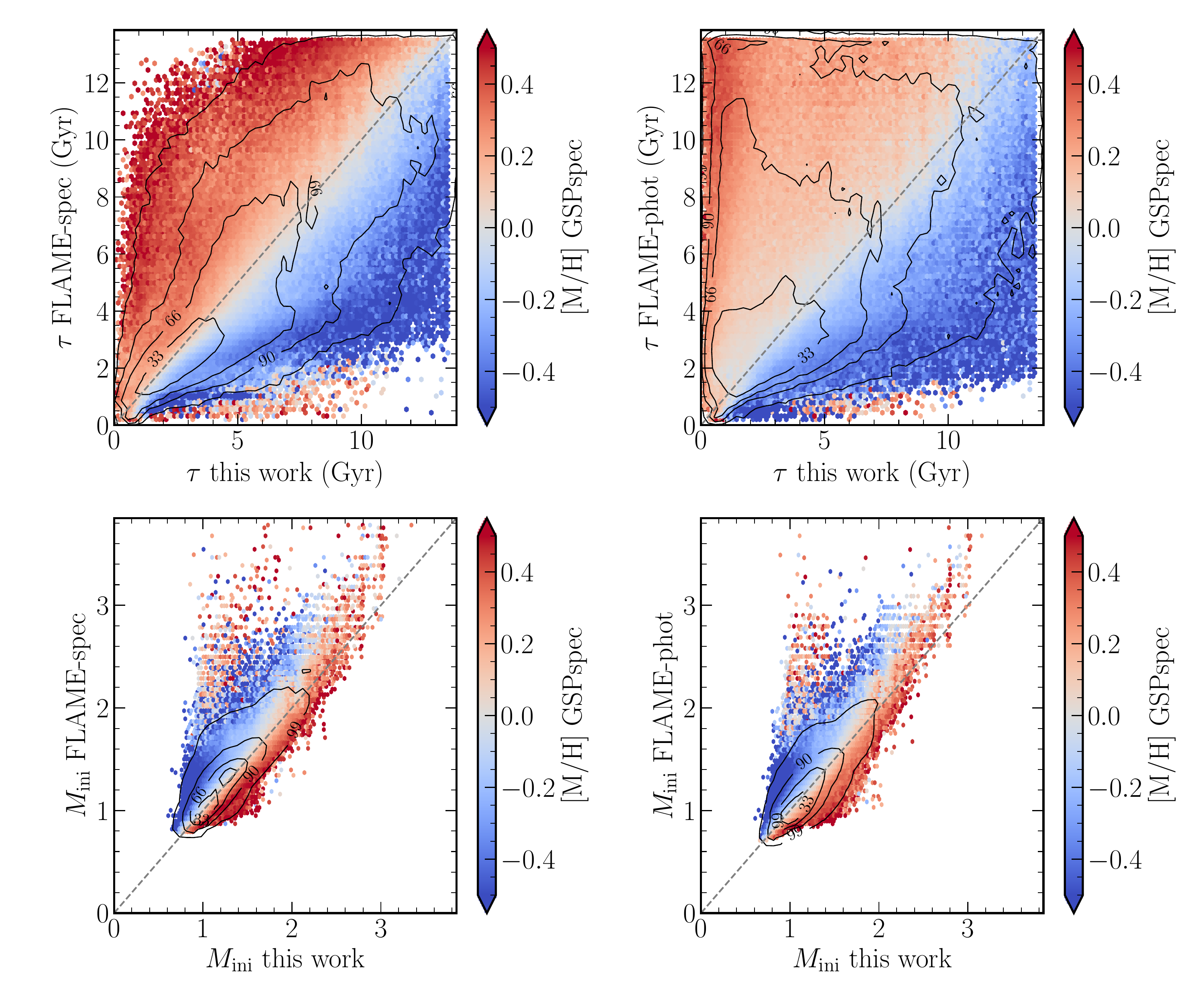}
\caption{Comparison of the ages (top) and the masses (bottom) between the estimations obtained in this work and  with the ones of FLAME-Spec (left) and FLAME-Phot (right). Only \gspspec~turn-off stars with all the first 13  \gspspec~flags equal to zero are selected here, with age uncertainties (both in our code and the FLAME modules) smaller than 50 per cent.  Contour lines contain 33, 66, 90 and 99 per cent of the distribution. The colour code is \gspspec's calibrated \meta.  }
\label{fig:FLAME_comparisons}
\end{center}
\end{figure}

Figure~\ref{fig:FLAME_comparisons}  compares our ages and masses with the ones derived from FLAME, obtained either with \gspphot~input (i.e.  FLAME-Phot) or with uncalibrated \gspspec~(i.e. FLAME-Spec). We recall that our ages and masses, regardless if photometry is taken into account or not, are always estimated with the \gspspec~\teff, and the calibrated \gspspec~\logg, and  \meta.  To obtain this plot, only turn-off stars are selected (based on calibrated \gspspec~parameters), since FLAME results are less reliable for the giants \citep[see][]{GaiaDR3-CU8-1, GaiaDR3-CU8-2}. We also imposed \gspspec~ parameter flags smaller or equal to 1 and relative age uncertainties (based on the different age estimations) smaller than 50 per cent.

The agreement of FLAME with our ages and masses can be considered as adequate only for metallicities close to solar values. 
Figure~\ref{fig:FLAME_comparisons} shows that as soon as \meta~is different than zero, then the disagreement becomes non negligible. Indeed, for stars with $\meta=\pm0.15\dex$, we find mean age offsets of $0.09\pm1.29\Gyr$ and $-0.01\pm 1.40\Gyr$ compared to FLAME-Spec and FLAME-Phot, respectively. These values  increase to $1.68\pm1.60\Gyr$ and $1.63\pm 1.76\Gyr$ when considering stars with $\meta<-0.15\dex$ and  
$2.96 \pm 1.82\Gyr$ and  $2.96\pm1.89\Gyr$ for stars with $\meta<-0.5\dex$. We therefore constrain the statement found in \citet{GaiaDR3-CU8-2} that suggest to treat with caution FLAME ages for stars with \meta$\leq-0.5\dex$, to an even narrower metallicity range of $\pm0.15\dex$ around solar values.

%%%%%%%%%%%%%%%%%%%%%%%%%%%%%%%%%%%
\subsection{Calculation of the reddening and the extinction}
\label{sec:extinctions}

To derive the extinctions, one must first compute the reddening $\EBPRP$, obtained by measuring the difference between the projected and the observed colours (\BP$-$\RP). 
Extinction, $A_G$, is then computed using:
\begin{equation}
%A_G=1.83037\cdot \EBPRP, 
A_G= c_\theta \cdot \EBPRP, 
\end{equation} 
where $c_\theta$ is a constant that depends on the  stellar type\footnote{For extinction converters as a function of the passband and the stellar atmospheric parameters see \url{https://www.cosmos.esa.int/web/gaia/dr3-astrophysical-parameter-inference}  }  \citep[][]{GaiaDR3-CU8-1, Fouesneau_dustapprox_2022}.

\begin{figure}[t!]
\begin{center}
\includegraphics[width=\linewidth, angle=0]{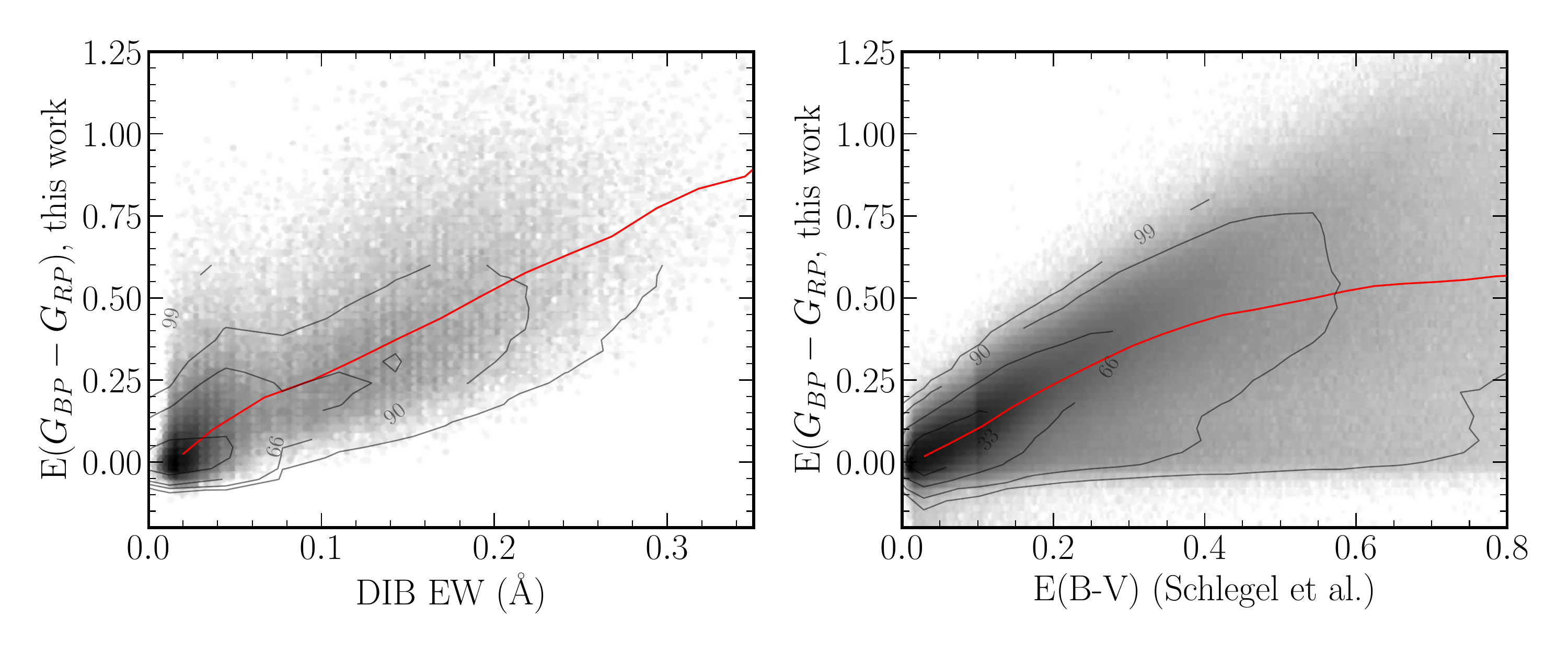}
\caption{Comparison of $\EBPRP$ derived in this work, with the equivalent widths of Diffuse interstellar bands at 862\,nm  derived from the same \gspspec~spectra (left) and the \citet{Schlegel98} $E(B-V)$ with the \citet{Bonifacio00} correction (right). For both plots, the grey colour-scale is the logarithm of the number of stars in one bin, and the red lines represent the running median. The correlations are, in both cases, overall very good.  }
\label{fig:EBPRP_DIB_SFD_comparisons}
\end{center}
\end{figure}

Left plot of Fig.~\ref{fig:EBPRP_DIB_SFD_comparisons} compares the reddening $\EBPRP$ derived in our work, with the equivalent widths of the diffuse interstellar band at $\sim862$\,nm, as derived in \citet{GaiaDR3_PVP_DIB}. The right plot compares with the  \citet{Schlegel98} $E(B-V)$ reddening (corrected as described in  Sect.~\ref{sec:ages_compilation} towards the regions with the largest extinction). Overall, a good correlation is found with both dust proxies, confirming our estimations of reddening. We note, however, that some of our targets have $\EBPRP$ values lower than zero. We decide not to artificially put them equal to zero, but draw the attention that the associated uncertainties of these stars should be considered.

\begin{figure}[t!]
\begin{center}
\includegraphics[width=\linewidth, angle=0]{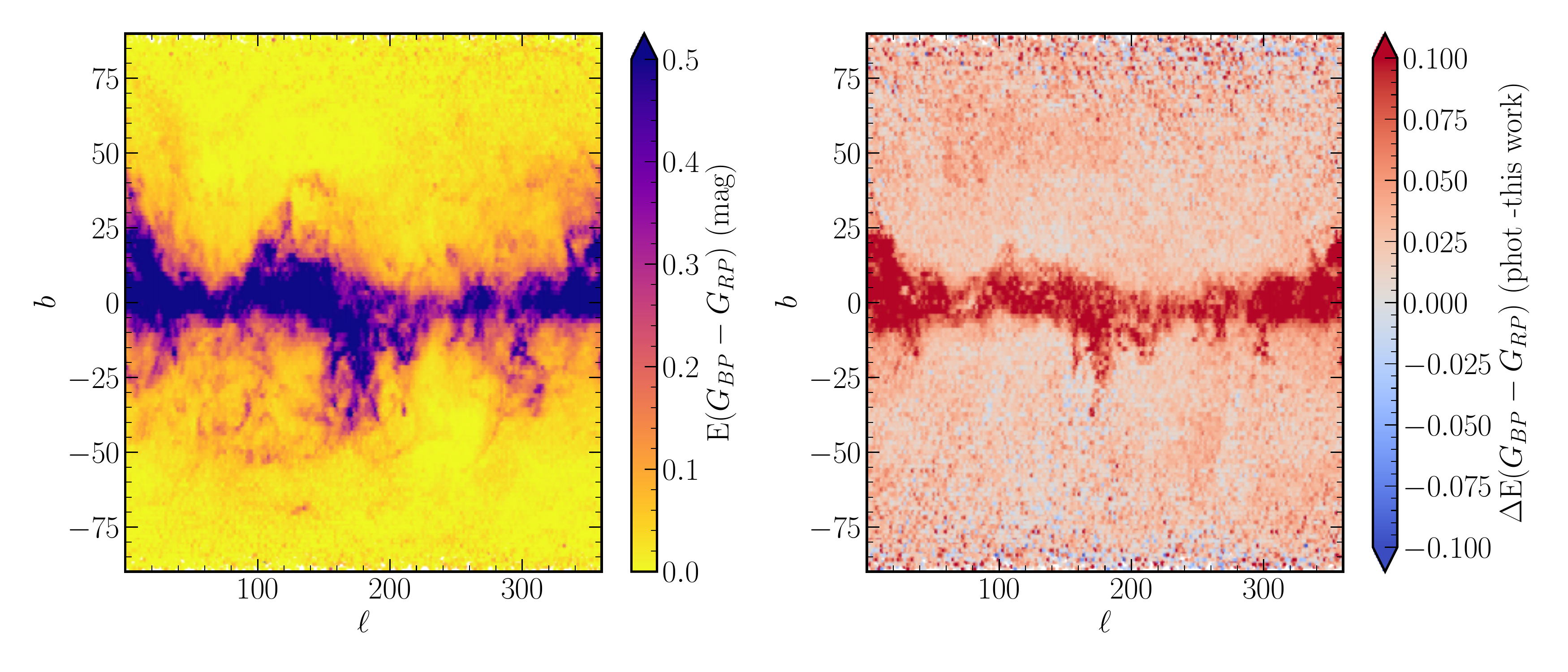}
\caption{Left: $\EBPRP$ distribution in galactic $(\ell,b)$ coordinates. Right: Residuals between the $\EBPRP$ estimation from \gspphot~and ours. }
\label{fig:EBPRP_galactic}
\end{center}
\end{figure}

Left plot of Fig.~\ref{fig:EBPRP_galactic} shows the reddening map derived  for the sample. 
The right plot of the same Fig. shows the residuals between  \gspphot's $\EBPRP$ values and ours. The median difference  above and below galactic latitudes of $|b|=20^\circ$ is 0.019\,mag and 0.052\,mag, respectively.  The agreement is rather good, acknowledging that  \gspphot~ does not use the same input data or parameters than we do, and that it is precisely towards highly reddened regions that degeneracies between  $E(BP-RP)$ and \teff~make the  \gspphot~ parameterisation challenging.

%%%%%%%%%%%%%%%%%%%%%%%
\subsection{Absolute magnitudes and line-of-sight distances}
\label{sec:Absmags}
\begin{figure}[t!]
\begin{center}
\includegraphics[width=\linewidth, angle=0]{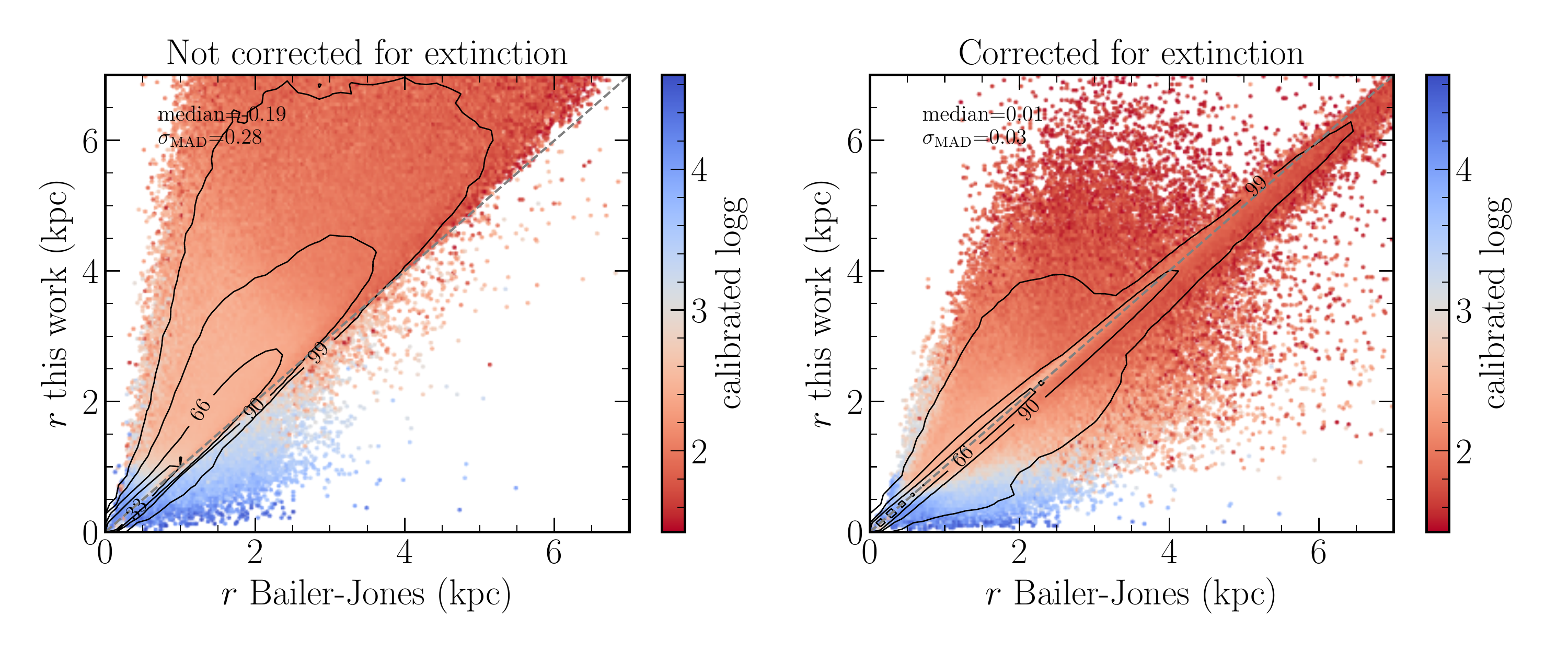}
\caption{Comparison of the line-of-sight distances derived from the projected $M_G$ and extinction, with the geometric ones from \citet{Bailer-Jones21}.  No cuts in the uncertainties on the \citet{Bailer-Jones21} distances or in our line-of-sight distances are made.  }
\label{fig:distance_comparisons}
\end{center}
\end{figure}

Finally, to validate more thoroughly the projected absolute magnitudes, we compare the geometric line-of-sight distances, $r$, of \citet{Bailer-Jones21} with the ones derived via the distance modulus (correcting  for the projected extinctions, assuming $c_\theta=1.83037$, 
valid for solar-type stars $\pm1500\,K$). Results are shown in Fig.~\ref{fig:distance_comparisons} (left: before correcting for the extinction, right: after applying the correction). They suggest that we find a very good agreement between the two distance estimations, with a null median residual and a dispersion of 20\,pc.  
We find that the 1~per~cent of  stars that have the largest disagreement with the \citet{Bailer-Jones21} distances have also either large differences between the input \logg~ and the projected one ($>0.3$) if these stars are dwarfs, or large age uncertainties (larger than 50 per cent) if these stars are giants.

Finally, we note that the very good agreement that is found is not necessarily a direct consequence of the use of $r$ when projecting the absolute magnitudes. Indeed,  we recall that when we project the absolute magnitudes, extinction is ignored. As it can be seen on the left plot of this Figure (that does not correct for the derived interstellar reddening), when the estimated $A_G$ is neglected when computing the distance, the agreement between the two distance estimates is rather poor and biased.

%%%%%%%%%%%%%%%%%%%%%%%%%%%%%%
\section{Age and mass correlations with the orbital parameters and the positions in the Galaxy}
\label{sec:positions_orbits_illustrations}
In this section we illustrate the quality and the limitations of our projected parameters. To do so, we first compute the orbital parameters for all of the stars with measured radial velocities from the RVS and available astrometry.   Then, we correlate them with the stellar ages and masses for a high-quality \gspspec~sample, requiring the first 12 quality flags of \gspspec~to be smaller or equal to one (except {\tt fluxNoise\_flag} that we require to be smaller or equal to 2), the {\tt KMgiantPar\_flag} that we require equal to zero, and the relative age uncertainty that we require to be smaller than 50 per cent. 

\subsection{Determination of Galactocentric positions, velocities and orbital parameters}
\label{sec:positions_orbits}
The Galactocentric positions $X,Y, Z$ (in cartesian coordinates), $R$ (Galactocentric cylindrical radius), and cylindrical velocities (radial $\vr$, azimuthal $\vphi$, vertical $\vz$)  in a right-handed frame have been computed for all of the stars that have a Gaia radial velocity measured ($\sim 33$\,million targets). In order to do so, we used the star's right ascension, declination, line-of-sight velocity, proper motions and \citet{Bailer-Jones21} EGDR3 geometric and photogeometric Bayesian line-of-sight distances (therefore leading to two sets of positions, velocities and orbits).
The assumed Solar position is  $(R, Z)_{\odot}=(8.249, 0.0208)$\,kpc \citep[][]{Gravity20, Bennett19} and  velocities $(\vr, \vphi, \vz)_\odot=(-9.5, 250.7, 8.56)\kms$ \citep[][]{Reid20, Gravity20}.

 To compute the orbital parameters (actions, eccentricities, apocentre, pericentre, maximum distance from the Galactic plane), we used the Stäckel fudge method \citep{Binney12a, Sanders16, Mackereth18} with the Galpy code \citep{Bovy15}, in combination with the axisymmetric potential of \citet{McMillan17} (adjusted to our adopted solar position and  Local Standard of rest's velocity). 
 The lower and upper confidence limits (corresponding to the 16th and the 84th percentile) were obtained by propagating the uncertainties  of the line-of-sight distance, line-of-sight velocity, and proper motions using 20 Monte-Carlo realisations.  No correlation between proper motions and distance uncertainties were taken into account, and we assumed a non-gaussian distribution for $r$, constructed as two half-Gaussians defined by the upper and lower confidence limits of $r$. 
 
 %%%%%%%%%%%%%%%%%%%%%
 \subsection{Galactic maps of ages and masses and identification of the spiral arms}
\begin{figure}[t]
\begin{center}
\includegraphics[width=\linewidth, angle=0]{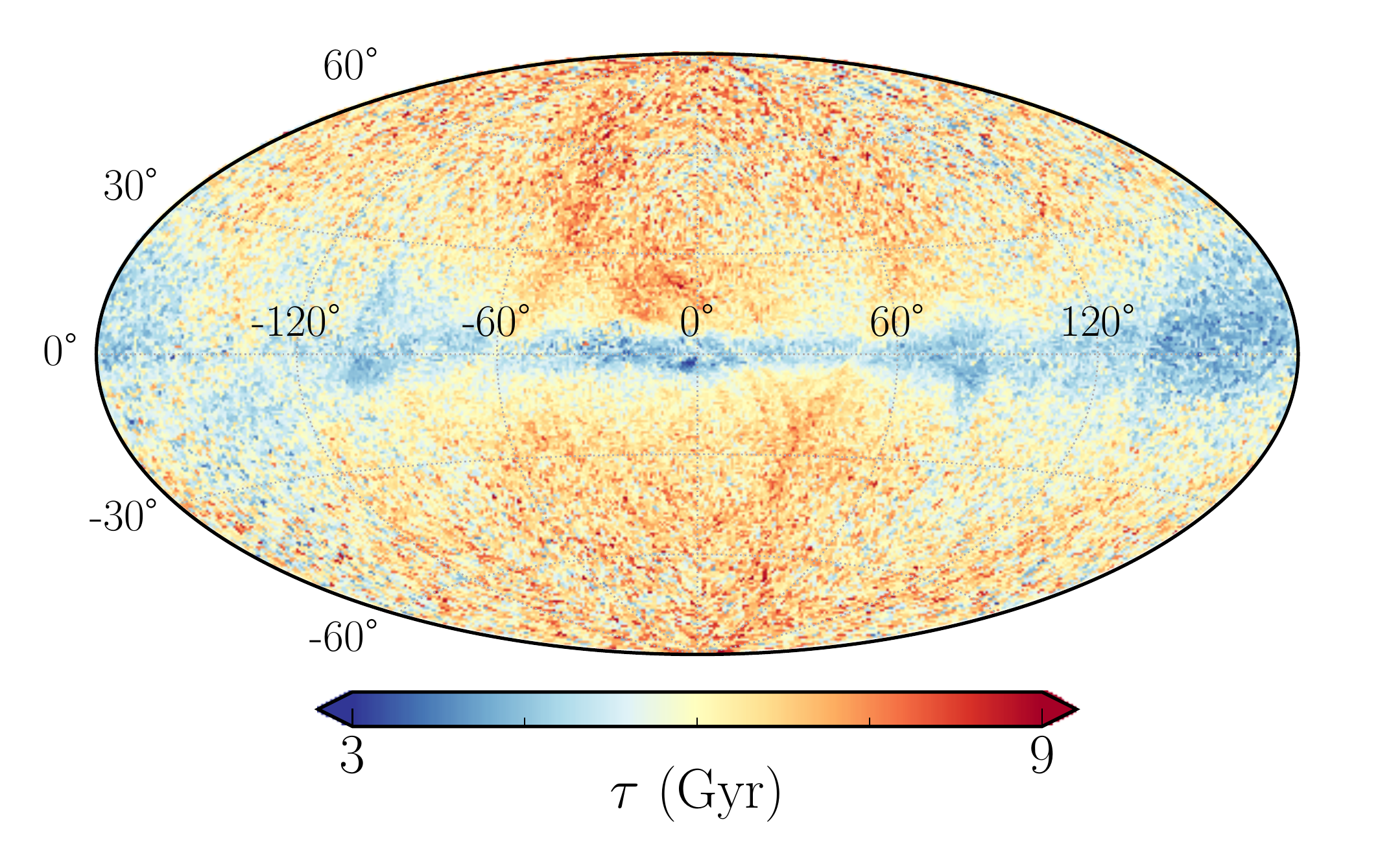}\\
\includegraphics[width=\linewidth, angle=0]{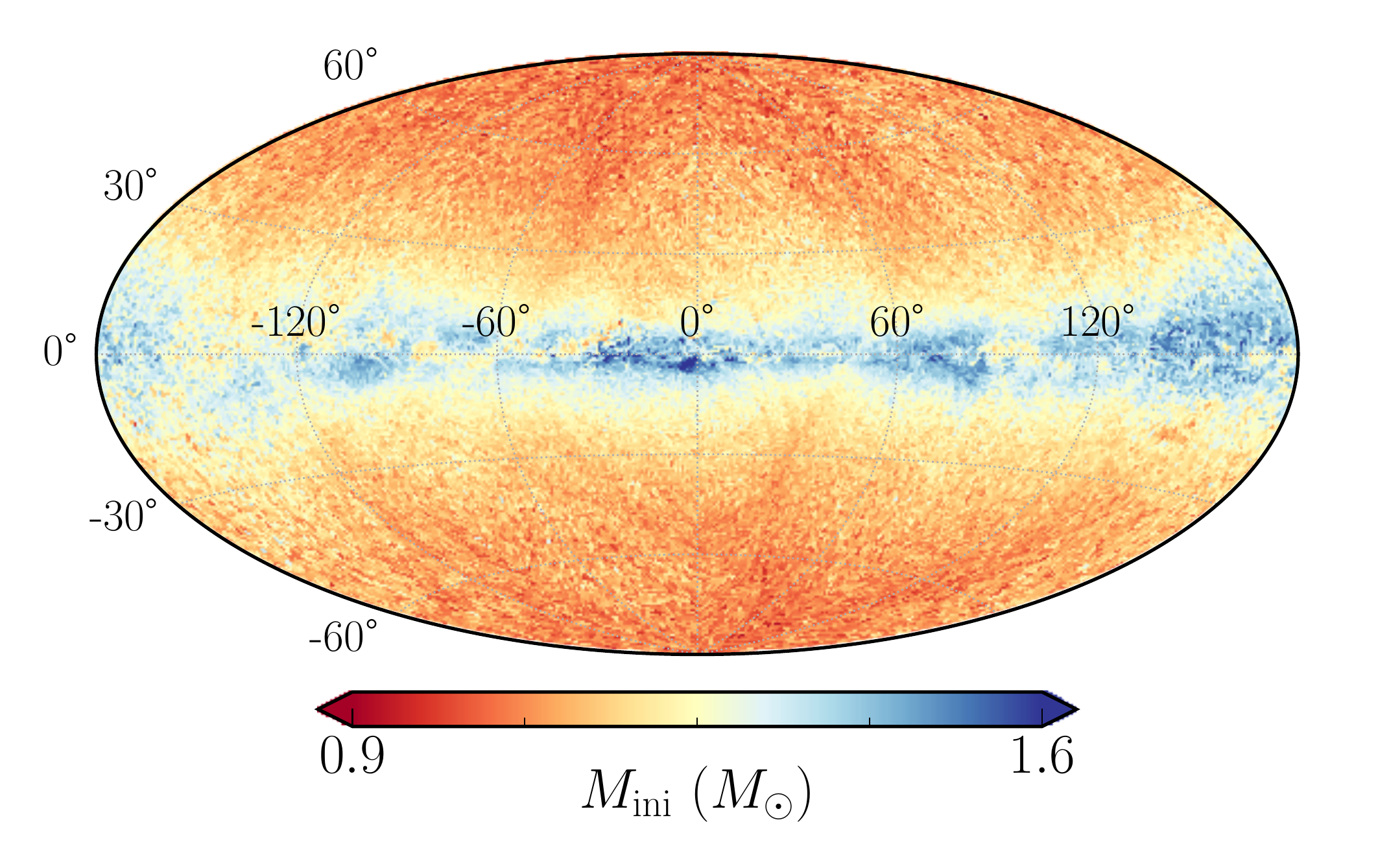}
\caption{Healpix projections (NSIDE=64) of the mean ages (top) and masses (bottom) for the compiled sample.    }
\label{fig:Aitoffs}
\end{center}
\end{figure}

Maps of the stellar ages and masses in Galactic Aitoff projection $(\ell,b)$ are shown in Fig.~\ref{fig:Aitoffs}.  Young and massive stars are predominantly found in (or close to) the Galactic plane,  within the regions of high reddening (see Fig.~\ref{fig:EBPRP_galactic}), as expected from the interstellar medium distribution  in the Milky Way \citep[e.g.][and references therein]{Kalberla09}. 

\begin{figure}[t!]
\begin{center}
\includegraphics[width=0.85\linewidth, angle=0]{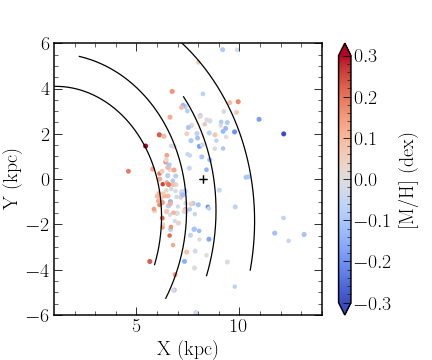}
\caption{Galactocentric cartesian XY projection of the position of the stars having $M_{\rm ini}\geq 4 M_\odot$, \logg$<2$ and $Z_{\rm max}<0.5\kpc$.  The Galactic  centre is located at $(X,Y)=0$ (at the left) and Galactic rotation is going clockwise. The colour code represents the metallicity of the stars, whereas the `+' sign is located at the Sun's position. The position of the Perseus, Local, Sagittarius and Scutum spiral arms, based on \citet{Castro-Ginard21} results,  are also plotted as black continuous lines.   }
\label{fig:massives}
\end{center}
\end{figure}

In Fig.~\ref{fig:massives} we plot the Galactocentric cartesian X-Y positions of stars that have an estimated initial mass greater than 4 solar masses, \logg$<2$ (to avoid massive  main sequence stars at the solar vicinity) and a maximum distance from the galactic plane during their orbit (i.e. $Z_{\rm max}$) less than 0.5\kpc. Superimposed, we also plot the position of the Perseus, Local, Sagittarius and Scutum spiral arms, based on \citet{Castro-Ginard21} analysis of open clusters with Gaia EDR3 data (adapted to match our assumed solar position).  The massive stars position follow very closely the one of the modelled spiral arms, consistent with the fact that the latter are regions where star formation takes place. Furthermore, one can also see the clear metallicity gradient within those stars, reflecting the metallicity of the interstellar medium at these locations (as these stars are found to be younger than 300 Myr). 

%Explaination for the spiral arms: The plots show the Perseus, Local, Sagittarius and Scotum arms in the X-Y plane, based on the parameters of \citet{Castro-Ginard21}, based on the EGDR3 data of open clusters. 
%   Based on Castro-Ginard et al., 2021 (EGDR3 and open clusters)
%   https://www.aanda.org/articles/aa/pdf/2021/08/aa39751-20.pdf
%   Note: Their Sun is located at R8.128kpc, Z=20.8pc
%

%%%%%%%%%%%%%%%%%%%%%%%%%%%
\subsection{Age-metallicity relation as a function of orbital actions}
\begin{figure*}[th!]
\begin{center}
\includegraphics[width=\linewidth, angle=0]{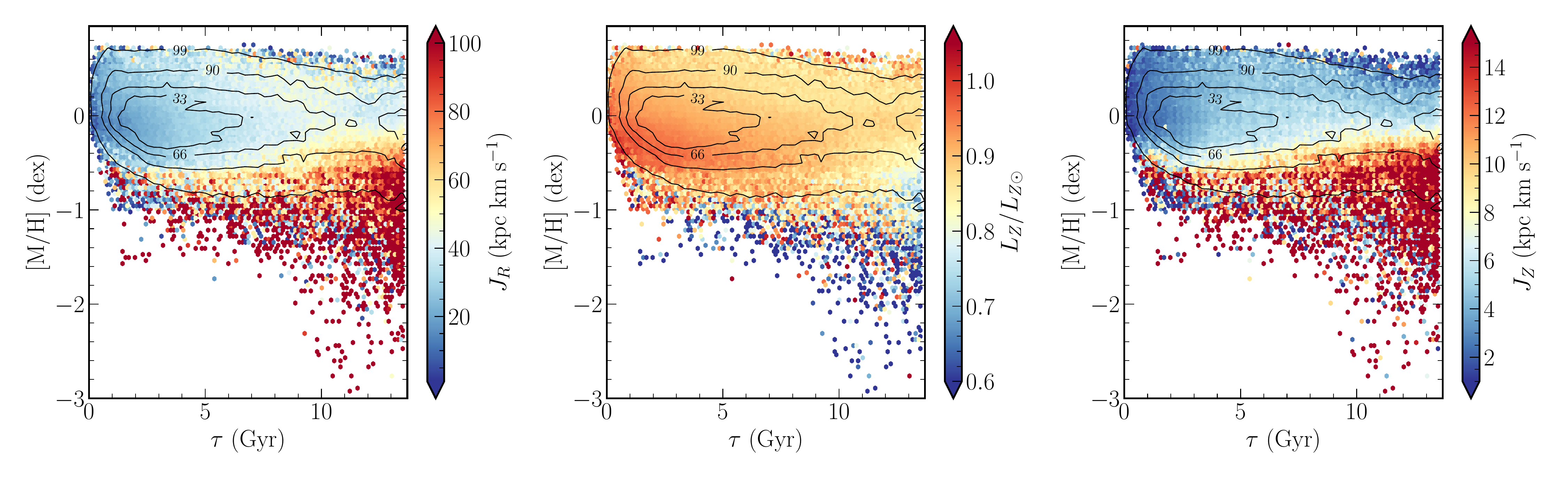}
\caption{Age-metallicity relations as a function of the radial and vertical actions ($J_R$, $J_Z$, first and third panel respectively) and normalised  angular momentum ($L_Z$ normalised by the Sun's value, middle panel), for stars closer than 1\kpc~ from the Sun.  }
\label{fig:Age_metallicity}
\end{center}
\end{figure*}

Figure~\ref{fig:Age_metallicity} shows the age metallicity-relation we derive for the stars within 1\kpc~from the Sun, colour-coded by the three different computed actions. Overall, we find a flat trend over the entire age-range, in agreement with previous studies \citep[e.g.][]{Edvardsson93, Casagrande11, Bergemann14, Feuillet19}. 
Young stars ($\tau\lesssim2\Gyr$) of sub-solar metallicity tend to have low $J_R$ and $L_Z/L_{Z\odot}$ above one, compatible with stars visiting the Solar neighbourhood on slightly eccentric orbits from the outer disc (we find that these stars have $e<0.1$).  

Super-Solar metallicity stars are found at all ages, with perhaps a slight decrease of their number for ages above $10\Gyr$ that may be due to our age prior. However, it is worth noticing that whereas the youngest super-solar metallicity stars  have low $J_R$ and $L_Z/L_{Z\odot}$ close to 0.9, the older ones have on average a larger radial action and more eccentric orbits ($e>0.3$) suggesting that they just visit the Solar neighbourhood, while being close to their apocentre. Interestingly, we also find old ($\tau>8\Gyr$) metal-rich stars with normalised  angular momentum around one and low values of radial and vertical actions. These stars are obvious candidates of objects  having experienced churning processes, i.e.  stars that moved  far from their birthplace without changing their eccentricity, via corotation resonances with the spiral arms or the Galactic bar,  \citep{Schonrich09a, Minchev13, Kordopatis15a}. 

Finally, we find that metal-poor stars ($\meta<-1\dex$) within 1\kpc~ from the Sun are predominantly old, with large radial and vertical actions and low angular momenta, as one would expect for typical halo stars.

%%%%%%%%%%%%%%%%%%%%%%%%%%%%
\subsection{Age-velocity dispersion and $Z_{\rm max}$-metallicity relations}
\begin{figure}[t!]
\begin{center}
\includegraphics[width=\linewidth, angle=0]{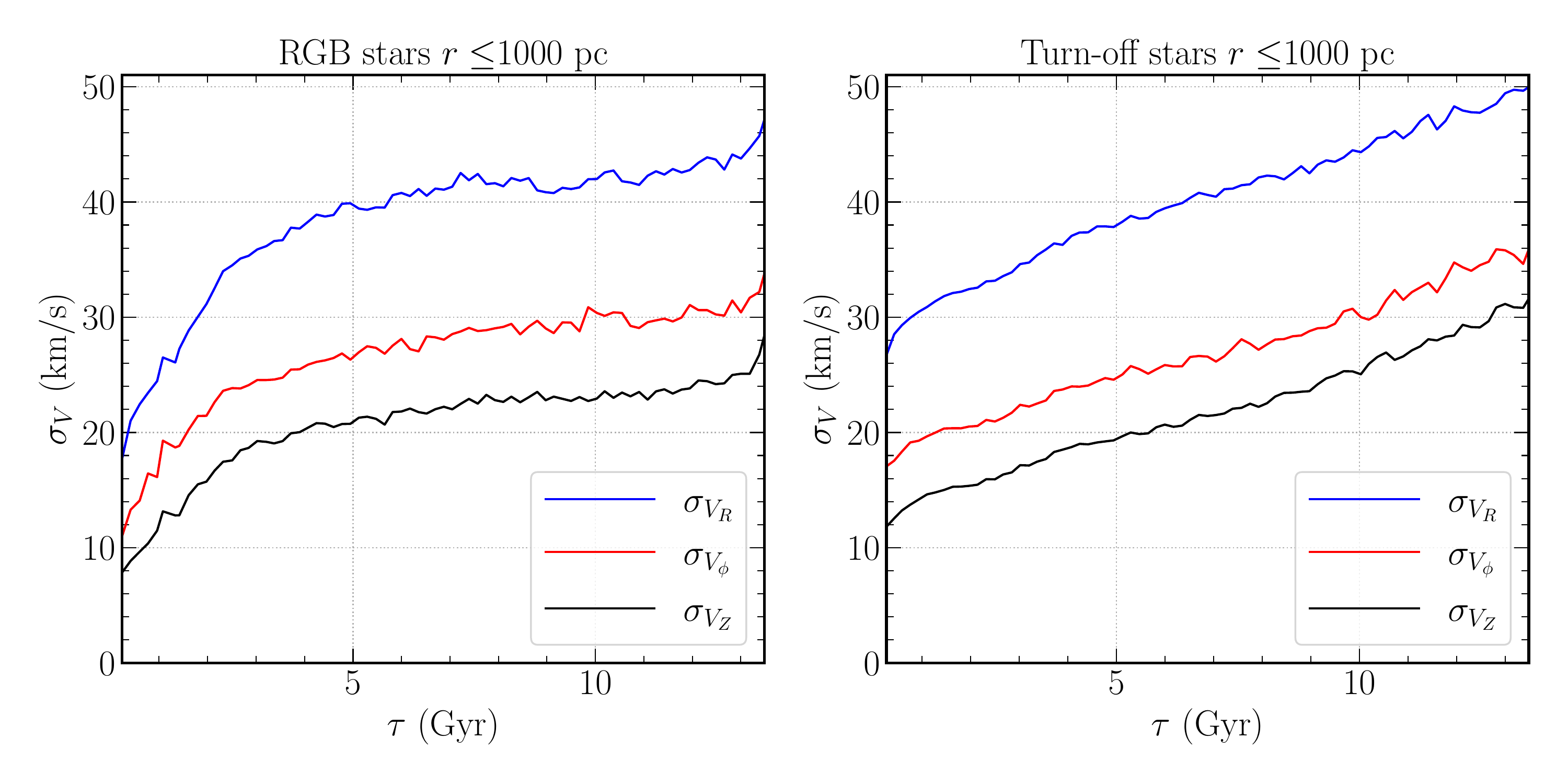}
\caption{Age-velocity dispersion for the RGB stars (left) and  for the turn-off stars (right) closer than $1\kpc$~from the Sun. A clear trend is found in both samples, for each velocity component. However, the fact that these trends do not exhibit the same shape, highlights the different precisions achieved for each stellar type: RGB stars tend to have under-estimated ages for old stars, whereas the younger TO stars tend to have over-estimated ages. }
\label{fig:Age_sigmavels}
\end{center}
\end{figure}

\begin{figure}[t!]
\begin{center}
\includegraphics[width=\linewidth, angle=0]{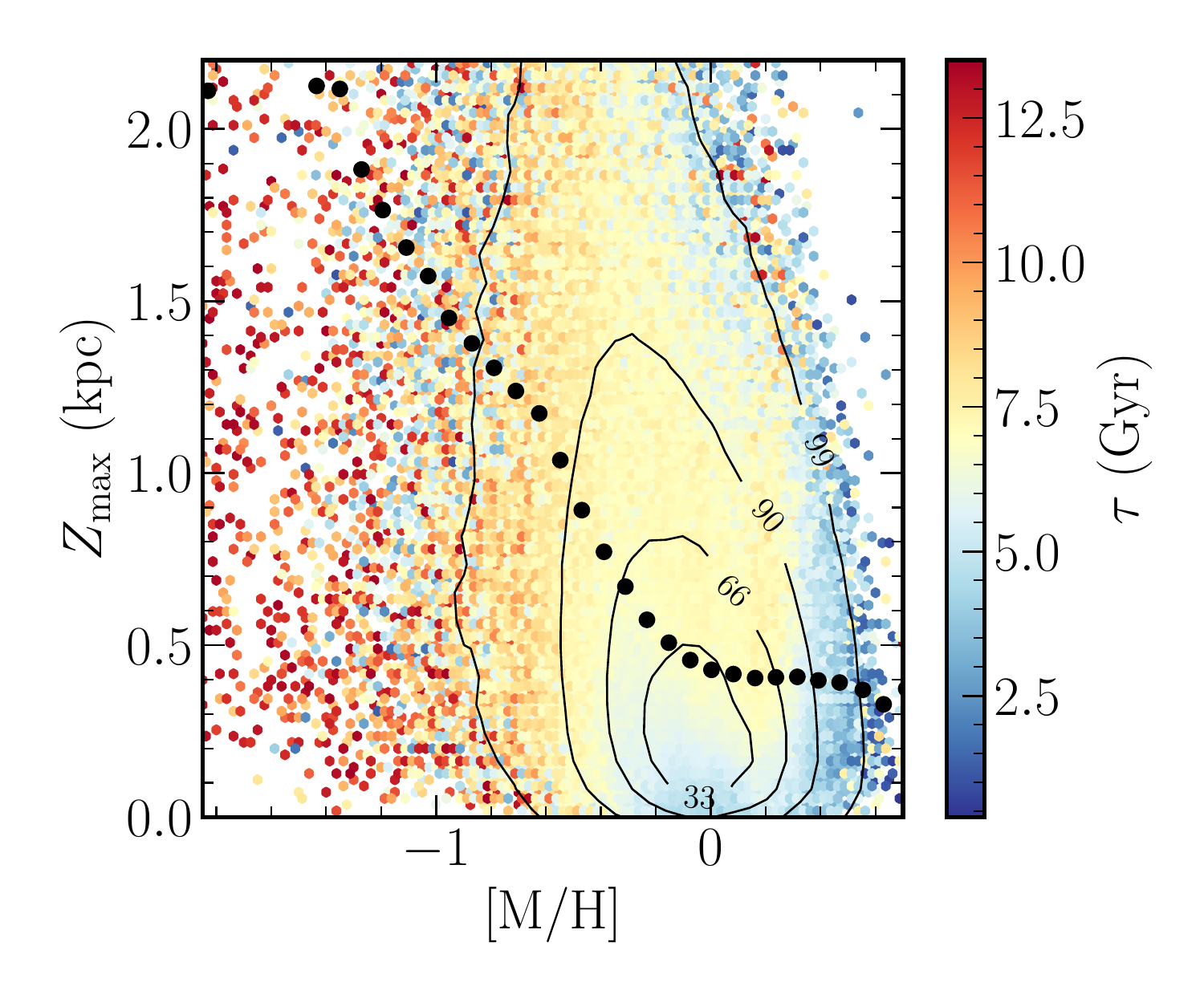}
\caption{Maximum distance from the Galactic plane reached during a star's orbit ($Z_{\rm max}$, in \kpc) versus metallicity, colour-coded by age (in Gyr). A clear increase of $Z_{\rm max}$ as a function of metallicity and age can be seen from the black dots on the figure that represent the running mean. This trend is  compatible with what one would expect for a transition between thin and thick discs and between thick disc and halo. }
\label{fig:Zmax_meta}
\end{center}
\end{figure}

Figure~\ref{fig:Age_sigmavels} shows the age-velocity dispersion relation of our RGB sample (left) and our  turn-off sample (right), for stars closer than $1\kpc$ from the Sun. In agreement with previous results \citep[e.g.][]{Aumer09}, we find a clear increase of the velocity dispersions with age,   which is even more pronounced when selecting only the turn-off stars. 
The trend for old stars is starker for the turn-off sample, whereas the giants seem to have a stronger trend at the young side. These different behaviours are in agreement with the ones found in Sect.~\ref{sec:age_mass_validation}, using the APOKASC-2 and \citet{Casagrande11} datasets, that suggest that old giants tend to have under-estimated ages, whereas very young turn-off stars tend to have over-estimated ages.

\begin{figure*}[t!]
\begin{center}
\includegraphics[width=\linewidth, angle=0]{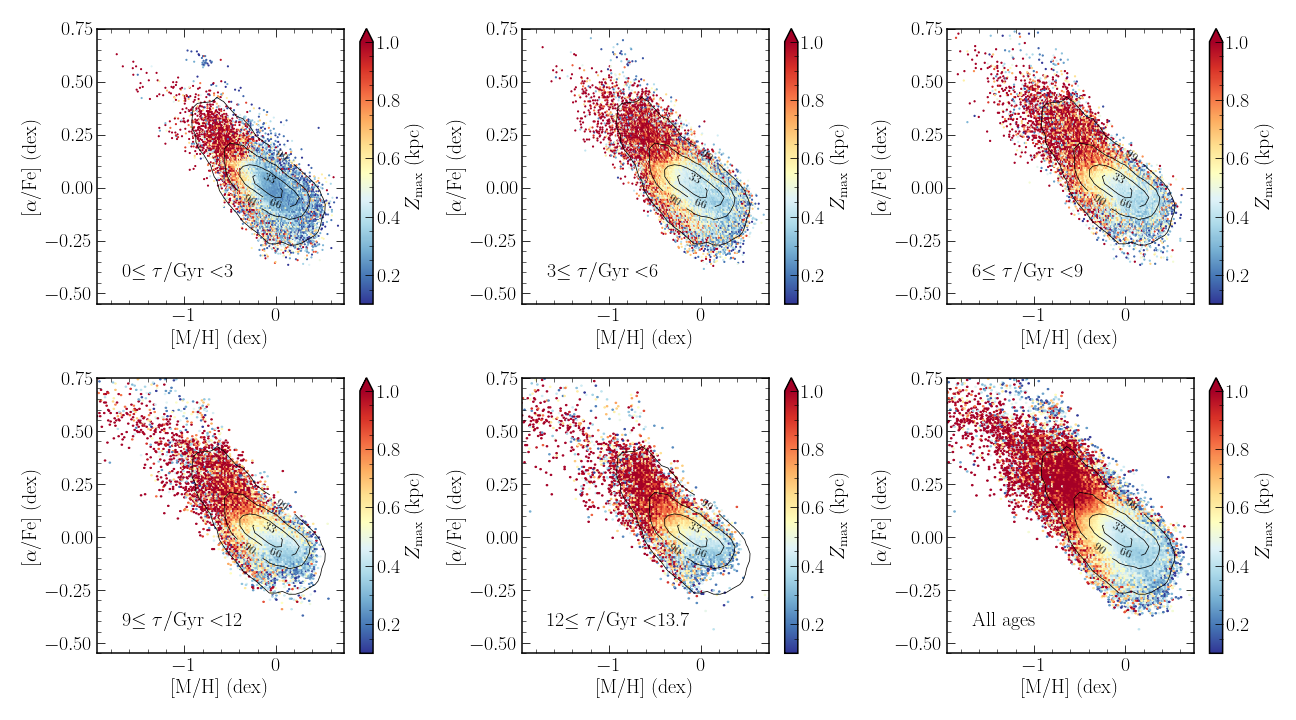}
\caption{Calibrated $\afe$ versus \meta, in $3\Gyr$ age-bins for giant stars (\logg$<3.5$) closer than $1.5\kpc$ from the Sun. The contour lines inside each panel have been evaluated for the sample considering all the age ranges, represented at the bottom-right panel. The colour-code corresponds to the maximum distance above the Galactic plane that a star can reach during its orbit.  }
\label{fig:alpha_meta_age}
\end{center}
\end{figure*}

Similarly, Fig.~\ref{fig:Zmax_meta} shows the maximum distance from the Galactic plane reached during a star's orbit  as a function of metallicity, colour-coded by age. Black circles represent the running mean in bins of \meta. It can be seen that stars with metallicities greater than $-0.2\dex$ tend to be young ($\tau<5\Gyr$) and confined in a thin disc configuration ($Z_{\rm max}<0.3\kpc$) whereas stars with $-0.2<\meta<-1\dex$ have $Z_{\rm max}<1.1\kpc$ and are globally younger than 10\Gyr, in agreement with the thick disc properties in the solar neighbourhood \citep[e.g.][]{Bensby14, Haywood18}.

%%%%%%%%%%%%%%%%%%%%%
\subsection{Chemical enrichment of the Galactic disc}

A different way to show the wealth of information in our sample is shown in Fig.~\ref{fig:alpha_meta_age}, where we plot the $\afe-\meta$ space, in various age-bins, as a function of the orbital parameters (in this case, $Z_{\rm max}$). The $\afe$ precision of \gspspec~ is not sufficient to separate the  well-known thin and thick disc chemical sequences \citep[see for example][obtained with high-resolution Gaia-ESO spectra]{Kordopatis15b}, nevertheless one can see that young stars have in majority low $\afe$ ratios ($<0.2$), high metallicity ($>-0.5$) and low $Z_{\rm max}$. As look-back time (i.e. age) increases, lower metallicity and higher $\afe$ regions get populated, with the low metallicity tail exhibiting the largest $Z_{\rm max}$ values.  
Chemical evolution models can then be fitted to these trends, in order to unfold the star formation history of Galaxy, together with its gas infall history. 

In addition to the inner evolution of the Milky Way, our dataset also allows to probe accreted populations present in the surveyed volume. For example, 
low-metallicity and low-$\afe$ stars with high $Z_{\rm max}$, associated to Gaia-Enceladus-Sausage \citep[][]{Belokurov18, Helmi18} stars are detected starting from $\tau>3\Gyr$ (with possible traces even below that we believe are sub-giants with under-estimated ages).
 Interestingly, the plot also shows some low-metallicity ($\meta<-0.6\dex$) $\alpha-$enhanced ($\afe>0.55\dex$) and low-$Z_{\rm max}$  ($<0.3\kpc$) stars present at all age bins. We find that these stars are in majority  targets with \teff$>6000$\,K and similar \logg$\sim3.5$ (also seen at the top left plot of Fig.~\ref{fig:Kiel_age_masses}). The true existence of these targets will have to be investigated further.

%%%%%%%%%%%%%%%%%%%%%%%%%%%%%%%%%%
\section{Conclusions} 
\label{sec:conclusions}

Using the calibrated atmospheric parameters  derived from Gaia spectra and the \gspspec~module, the photometry from 2MASS and Gaia-EDR3, and Bayesian line-of-sight distances estimated using Gaia-EDR3 parallaxes, we derived ages, initial masses and absolute magnitudes for $\sim$ 5 million targets in four different ways, depending on different combinations of parameters to project on isochrones. We propose a way to combine these different estimations, and publish a compiled catalogue of best projected parameters and their uncertainties.

We note that the reliability of the projected parameters is closely related to the one of the input data and their associated uncertainties. 
Indeed, biases in \teff, \logg, \meta~or distance modulus,  and/or underestimated errors on them,  may lead (depending on the stellar evolutionary phase) to biases on the output ages and masses. 
For this reason, a careful consideration of \gspspec's flags on the atmospheric parameters is necessary, according to the user's objectives,  in order to chose the parameters with the desired reliability (see Table. 2 of \citealt{GaiaDR3-GSPspec}).

Tests made comparing our ages and masses with reference catalogues of field stars, open clusters and globular clusters suggest that our code performs well, provided a filtering on the estimated relative age uncertainty (that we suggest $<$ 50 per cent), except for the older giants, for which ages tend to be under-estimated.
Ages are found to be the most reliable for turn-off stars, even when the \gspspec~parameters have large uncertainties, whereas age estimations for giants and main-sequence stars are also retrieved reliably (with uncertainties of the order of 2\Gyr) provided the extinction towards the star's line-of sight is smaller than $A_V\lesssim2.5$\,mag. Initial stellar masses are retrieved robustly for main-sequence and turn-off stars (dispersions compared to literature of the order of $0.1\,M_\odot$), whereas a filtering based on the age uncertainty of the giants is necessary to get reliable masses for the latter (dispersions compared to literature of the order of $0.3 M_\odot$).

We complete our catalogue with Galactocentric positions and velocities as well as orbital parameters (actions, eccentricities, apocenters, pericenters, maximum distance from the Galactic plane)  evaluated for the entire RVS sample, using an axisymmetric Galactic potential and commonly used orbital derivation methods and codes.
The catalogue, publicly available, is described in Table~\ref{tab:catalogue}.

%%%%%%%%%%%%%%%%%%%%%%%%%%%%%%%%%%%%%%%%%
\section*{Acknowledgments}
M. Fouesneau is warmly thanked for valuable comments, as well as CU8 and CU9 members that helped validating the stellar parameters used in this paper.
This work has made use of data from the European Space Agency (ESA) mission
{\it Gaia} (\url{https://www.cosmos.esa.int/gaia}), processed by the {\it Gaia}
Data Processing and Analysis Consortium (DPAC,
\url{https://www.cosmos.esa.int/web/gaia/dpac/consortium}). Funding for the DPAC
has been provided by national institutions, in particular the institutions
participating in the {\it Gaia} Multilateral Agreement.
This research made use of Astropy,\footnote{\url{http://www.astropy.org}} a community-developed core Python package for Astronomy \citep{astropy2013, astropy2018}. 
 The Centre National de la Recherche Scientifique (CNRS) and its SNO Gaia of the Institut des Sciences de l’Univers (INSU), its Programmes Nationaux: Cosmologie et Galaxies (PNCG) are thanked for their valuable financial support. 
ARB, GK and ES acknowledge funding from the European Union’s Horizon 2020 research and innovation program under SPACE-H2020 grant agreement number 101004214 (EXPLORE project). PM acknowledges support from the Swedish National Space Agency (SNSA) under grant 20/136.

\bibliographystyle{../aa}
\def\aj{AJ}\def\apj{ApJ}\def\apjl{ApJL}\def\araa{ARA\&A}\def\apss{Ap\&SS}
\def\mnras{MNRAS}\def\aap{A\&A}\def\nat{Nature}
\def\nar{New Astron. Rev.}

\bibliography{../master_bib}

\clearpage
\onecolumn

\begin{appendix}
\section{Output catalogue format}
%Table description
\begin{table*}[!h]
\caption{Description of the columns of the published catalogue}
\begin{center}
\begin{tabular}{ccc}
Name of column & Description & Units \\ \hline
source\_id & Gaia DR3 source ID & -- \\
age & Inferred age & Gyr\\
age\_error &  Uncertainty on the inferred age & Gyr\\
m\_ini & Inferred initial stellar mass & $M_\odot$\\
m\_ini\_error & Inferred uncertainty on the initial stellar mass & $M_\odot$\\
teff & Adopted projected \teff & K \\
logg & Adopted projected \logg & \dex \\
meta & Adopted projected \meta & \dex \\
p\_flavour & Adopted projection flavour for inferred ages and masses & \\
%A\_g & Inferred extinction using the projected \BP~and \RP & mag\\
%A\_g\_error &  Uncertainty on the inferred extinction & mag \\
G & Inferred absolute G magnitude & mag \\
G\_BP & Inferred absolute \BP~magnitude & mag \\
G\_RP & Inferred absolute \RP~magnitude & mag \\
ebprp & Inferred reddening using the projected \BP~and \RP & mag \\
ebprp\_error &  Uncertainty on the inferred reddening & mag \\ 
**$\_$spec & Parameters adopting the calibrated \gspspec~parameters only\\
**$\_$speck & Parameters adopting the calibrated \gspspec~parameters and the $K_s$ absolute magnitude\\
**$\_$specjhk  & Parameters adopting the calibrated \gspspec~parameters and the $J, H, K_s$ absolute magnitudes \\
**$\_$specjhkg  & Parameters adopting the calibrated \gspspec~parameters and the $J, H, K_s, G$ absolute magnitudes \\
\hline
x\_med\_dgeo & Median galactocentric Cartesian X position & \kpc \\
y\_med\_dgeo  & Median galactocentric Cartesian Y position & \kpc \\
z\_med\_\_dgeo  & Median galactocentric Cartesian Z position & \kpc \\
r\_med\_dgeo  & Median heliocentric line-of-sight distance with geometric prior from \citet{Bailer-Jones21} & pc \\
vr\_med\_dgeo &  Median galactocentric radial velocity obtained using r\_med\_dgeo & ${\rm \kms}$  \\
vphi\_med\_dgeo &  Median galactocentric azimuthal velocity obtained using r\_med\_dgeo & $\kms$ \\
vz\_med\_dgeo &  Median galactocentric vertical velocity obtained using r\_med\_dgeo & $\kms$\\
jr\_med\_dgeo & Median radial action obtained using r\_med\_dgeo &$\kpc\kms$  \\
jphi\_med\_dgeo& Median angular momentum (i.e. azimuthal action) obtained using r\_med\_dgeo&$\kpc\kms$  \\
jz\_med\_dgeo  & Median vertical action obtained using r\_med\_dgeo &$\kpc\kms$  \\
zmax\_med\_dgeo & Median maximum distance from the galactic plane obtained using r\_med\_dgeo  & \kpc\\
rapo\_med\_dgeo & Median apogalactic radius reached by the star, obtained using r\_med\_dgeo  & \kpc\\
rperi\_med\_dgeo & Median perigalactic radius reached by the star, obtained using r\_med\_dgeo  & \kpc\\
e\_med\_dgeo & Median eccentricity, obtained using r\_med\_dgeo  & \kpc\\
**\_upper\_dgeo & Upper confidence limit of the parameters \\ 
**\_lower\_dgeo  & Lower confidence limit of the parameters \\ 
**\_dphotogeo & Parameters obtained using the photogeometric distances from \citet{Bailer-Jones21}  \\ 
\hline

\end{tabular}
\end{center}
\label{tab:catalogue}
\end{table*}%

\section{Results of the isochrone projection with extinction}
\label{appendix:projection_extinction}
\begin{figure*}[ht!]
\begin{center}
\includegraphics[width=\linewidth, angle=0]{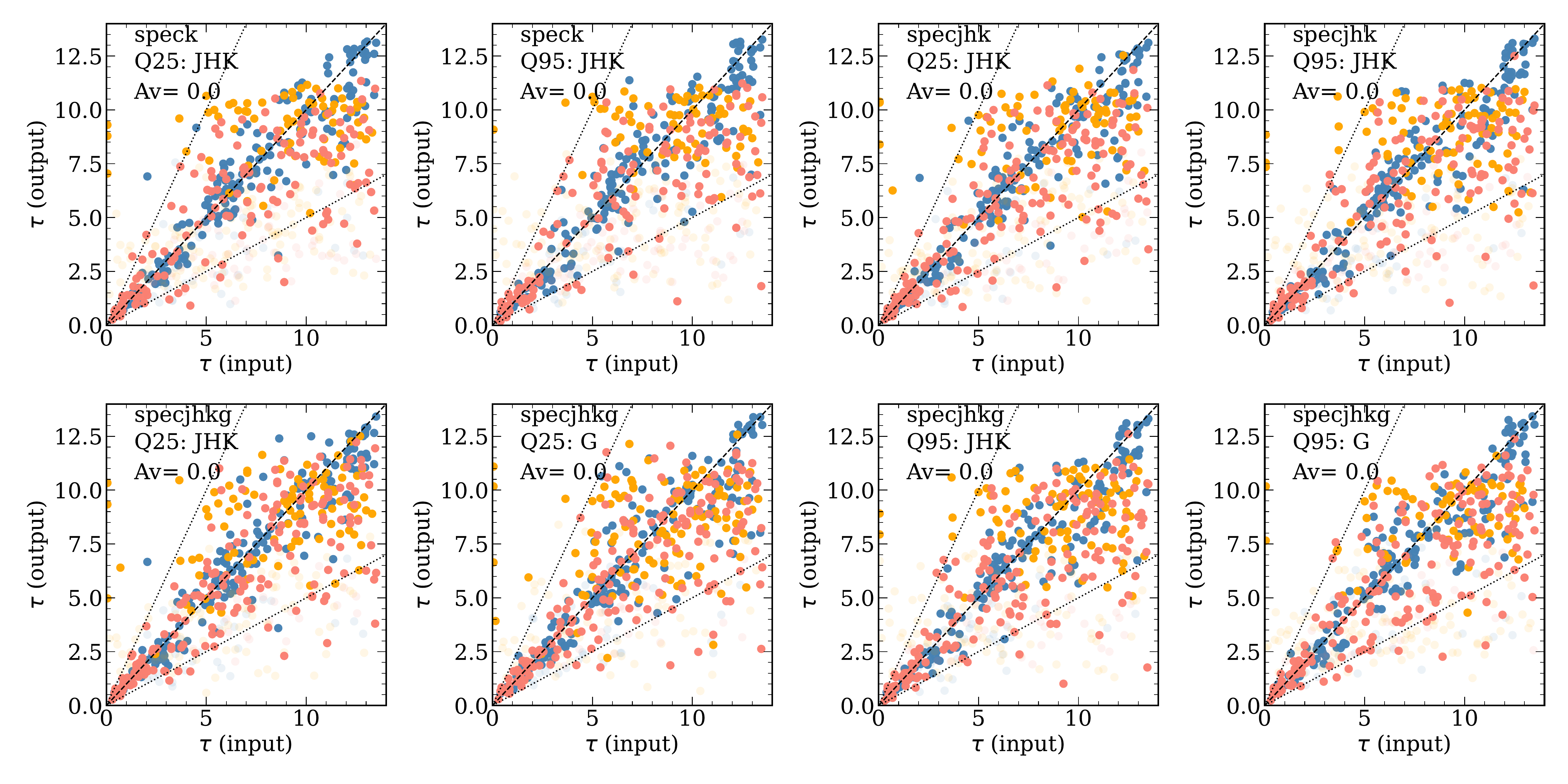}
\caption{Comparison of the input ages versus the output ones, for the  {\tt speck}, {\tt specjhk} and {\tt specjhkg} projections, with Q50 uncertainties in \teff, \logg~ and \meta, and uncertainties in $J,H,K_s$ and $G$ as indicated at the top left corner of each plot. No interstellar extinction is considered here. 
}
\label{fig:Synthetic_projection_ages_Av0.0}
\end{center}
\end{figure*}

\begin{figure*}[ht!]
\begin{center}
\includegraphics[width=\linewidth, angle=0]{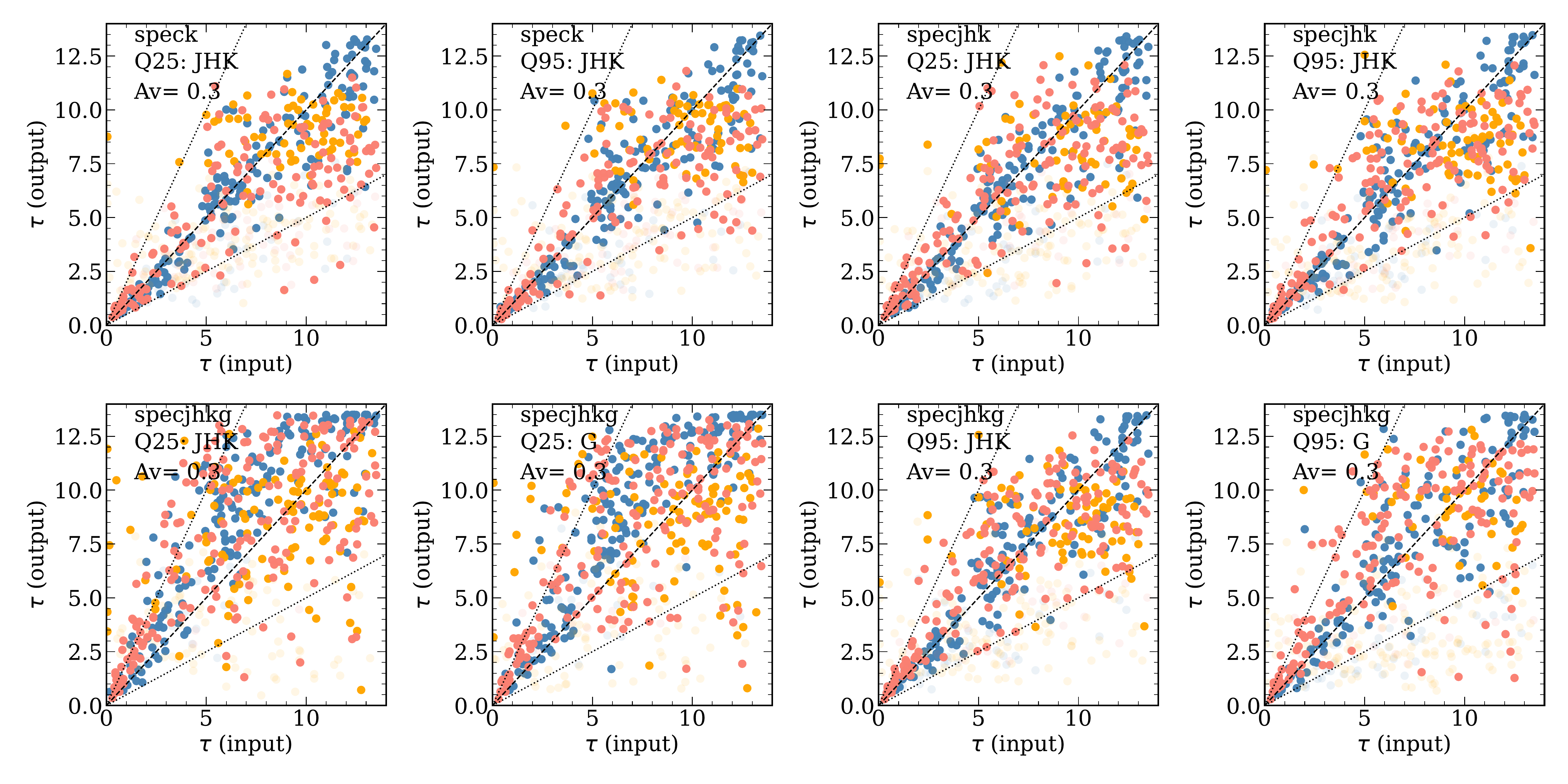}
\caption{Similar to Fig.~\ref{fig:Synthetic_projection_ages_Av0.0} with an extinction of $A_V=0.3$\,mag. }
\label{fig:Synthetic_projection_ages_Av0.3}
\end{center}
\end{figure*}

\begin{figure*}[ht!]
\begin{center}
\includegraphics[width=\linewidth, angle=0]{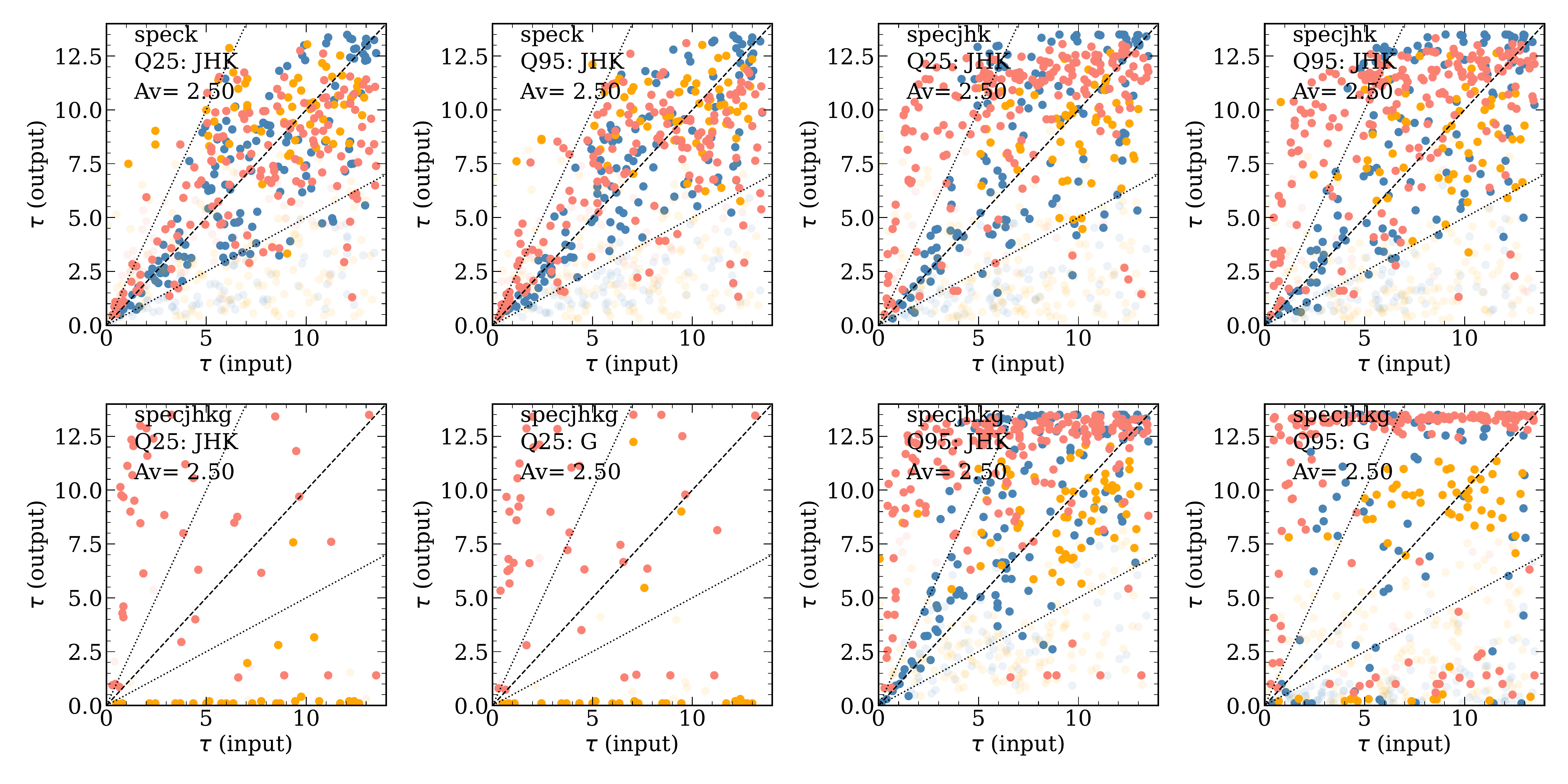}
\caption{Similar to Fig.~\ref{fig:Synthetic_projection_ages_Av0.0} with an extinction of $A_V=2.5$\,mag. }
\label{fig:Synthetic_projection_ages_Av2.5}
\end{center}
\end{figure*}

Figures~\ref{fig:Synthetic_projection_ages_Av0.0} to \ref{fig:Synthetic_projection_ages_Av2.5} are similar to Fig.~\ref{fig:Synthetic_projection_ages}. They consider only the {\tt speck}, {\tt specjhk} and {\tt specjhkg} projections, with Q50 uncertainties in \teff, \logg~ and \meta, and uncertainties in $J,H,K_s$ and $G$ as indicated at the top left corner of each plot. The input magnitudes are reddened according to the extinction $A_V$ labelled within each plot. 

\section{Open and globular cluster plots}
\begin{figure}[ht!]
\begin{center}
% WORKS GOOD: 
\includegraphics[width=0.75\linewidth, angle=0]{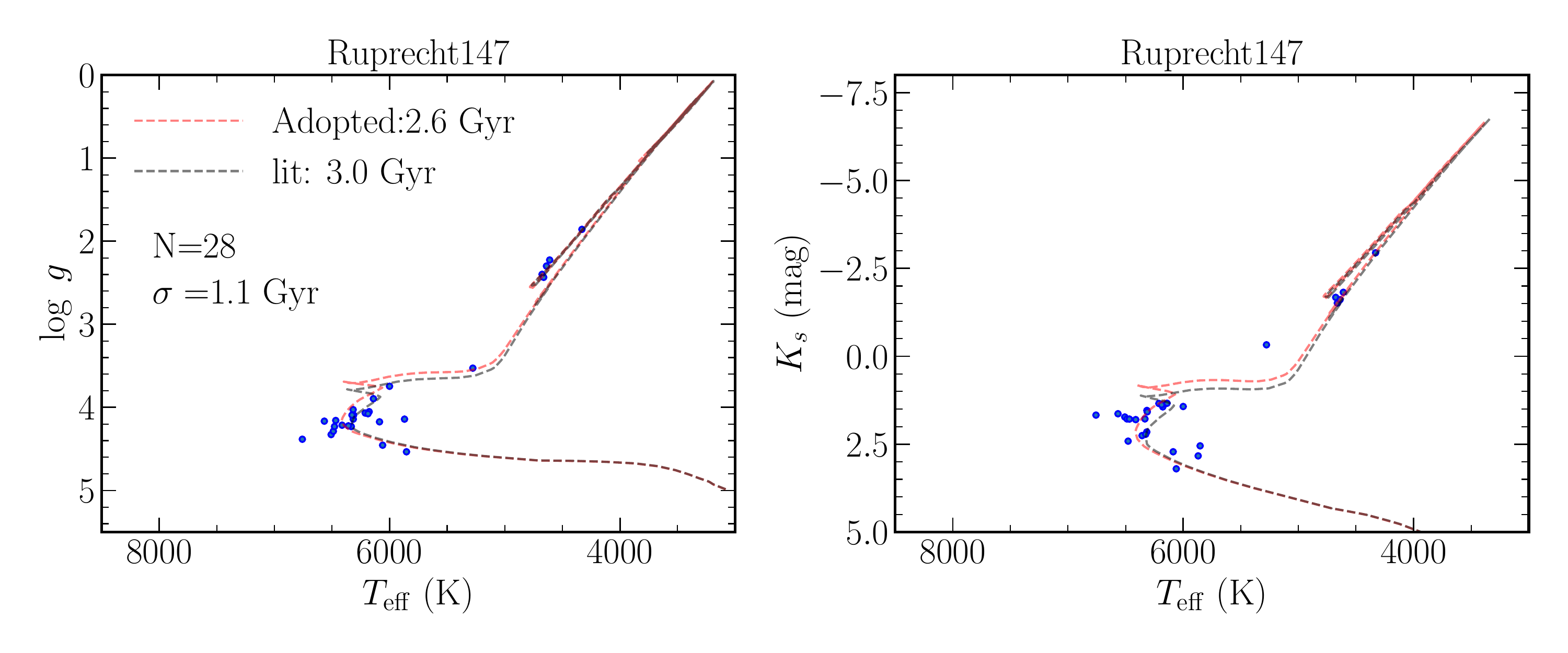}\\
\includegraphics[width=0.75\linewidth, angle=0]{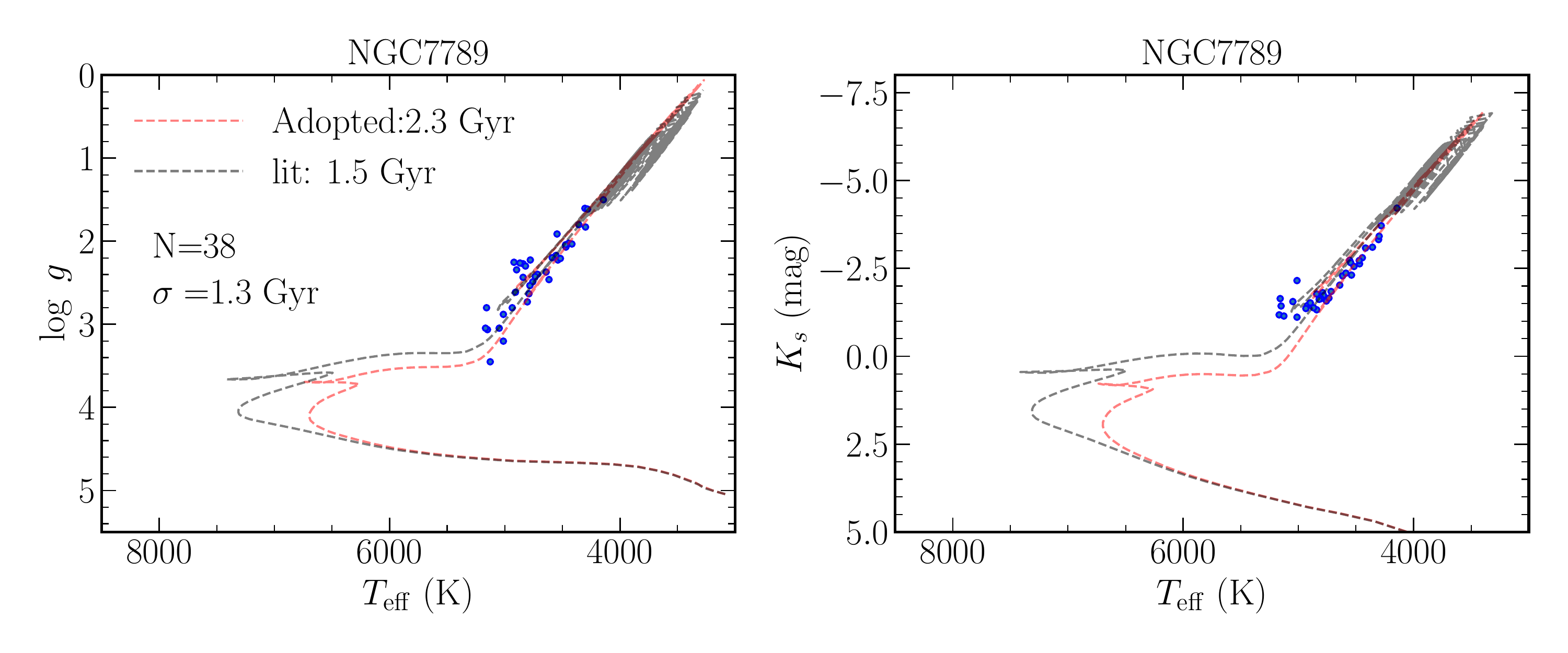}\\
% DOES NOT WORK GOOD
\includegraphics[width=0.75\linewidth, angle=0]{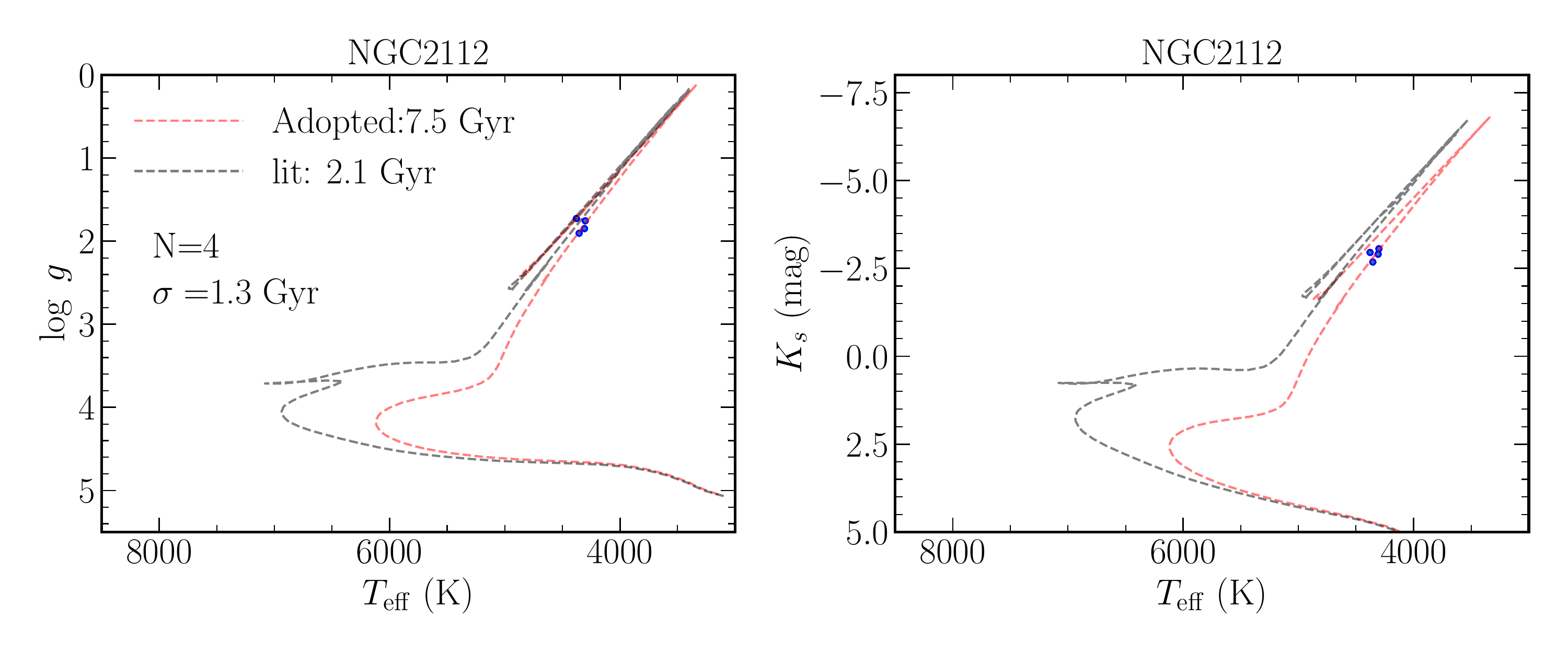}\\
\includegraphics[width=0.75\linewidth, angle=0]{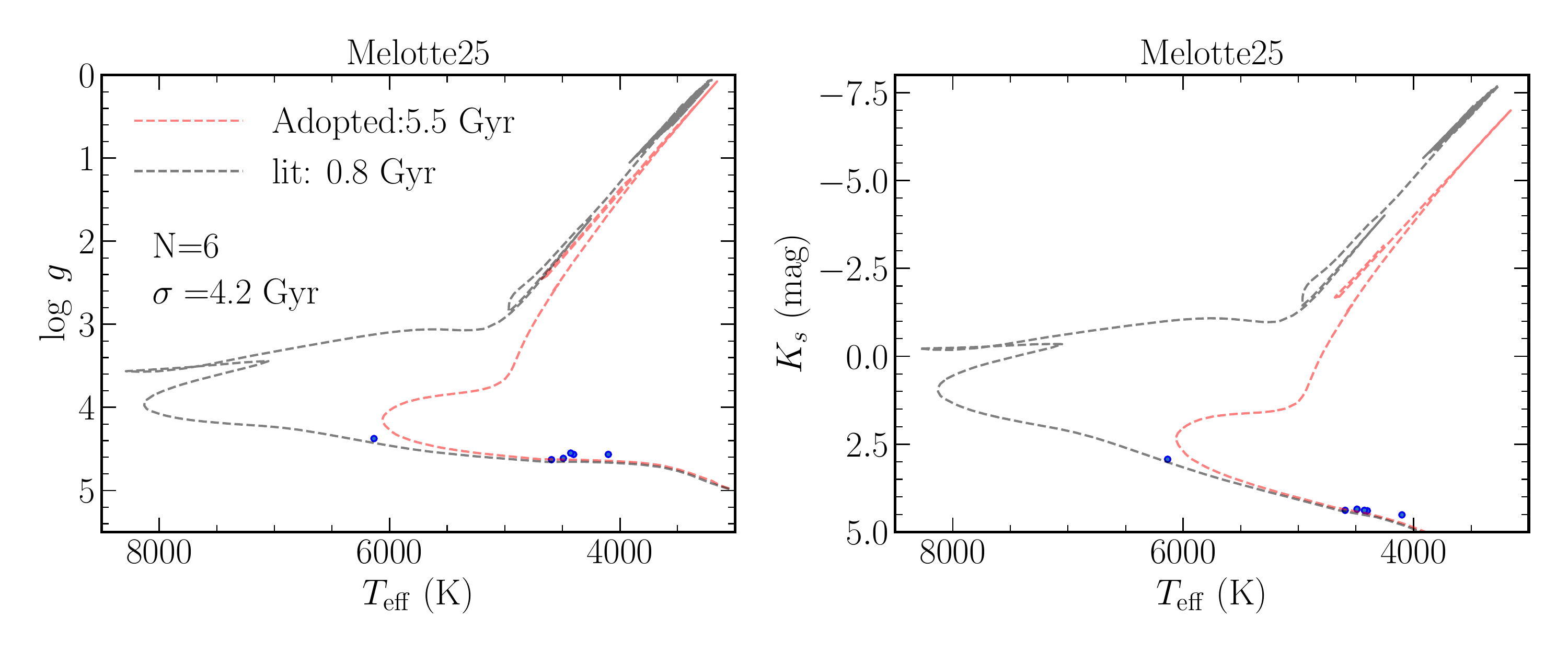}
\caption{\teff-\logg~(left) and \teff-$M_{\rm K_s}$ (right) diagrams for few cherry-picked open clusters.%, which names are written at the top of each plot. 
The isochrones of the adopted mean age (in red) and the reference age from \citet{Cantat-Gaudin20} are plotted in red and black, respectively, for the mean metallicity of the cluster as derived from the calibrated \gspspec~values. 
The first two rows show examples for which  our  ages and the literature ones are in good agreement. 
The bottom two rows show the opposite:  for NGC2112  our solution fits better the data than the younger isochrone, pointing towards offsets in the input \teff, \meta, or distance modulus. For NGC7789, our solution is not reliable,  since  main-sequence stars do not constrain enough the ages. }
\label{fig:cluster_comparisons_isochrones}
\end{center}
\end{figure}

\begin{figure}[ht!]
\begin{center}
\includegraphics[width=0.75\linewidth, angle=0]{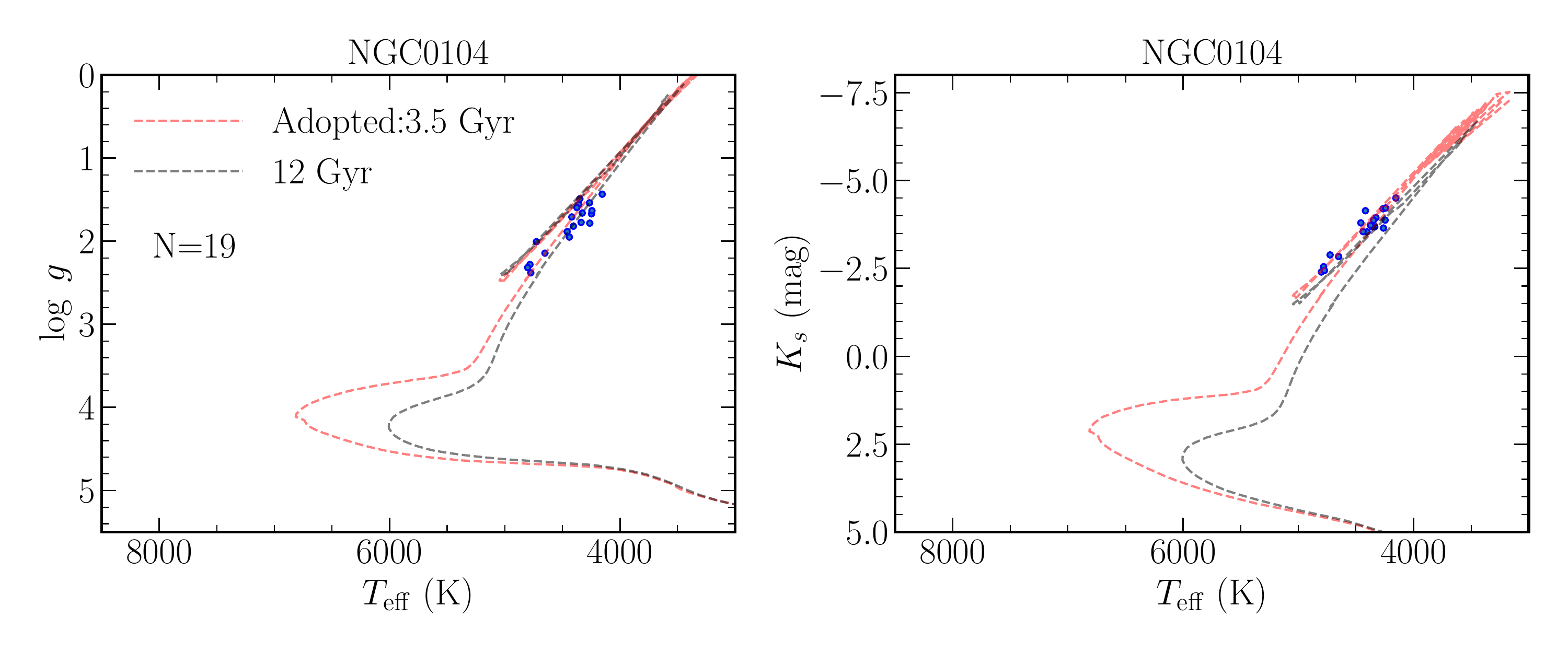}\\
\includegraphics[width=0.75\linewidth, angle=0]{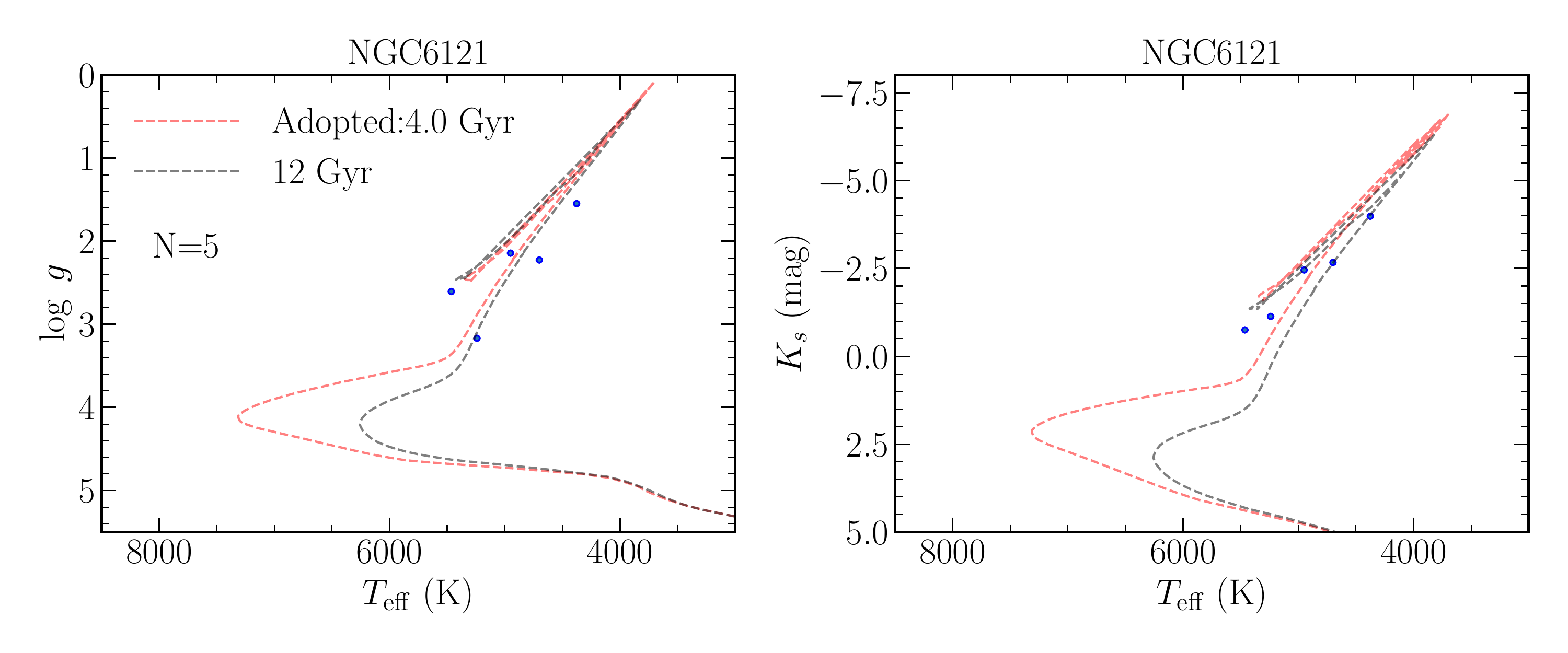}\\
\includegraphics[width=0.75\linewidth, angle=0]{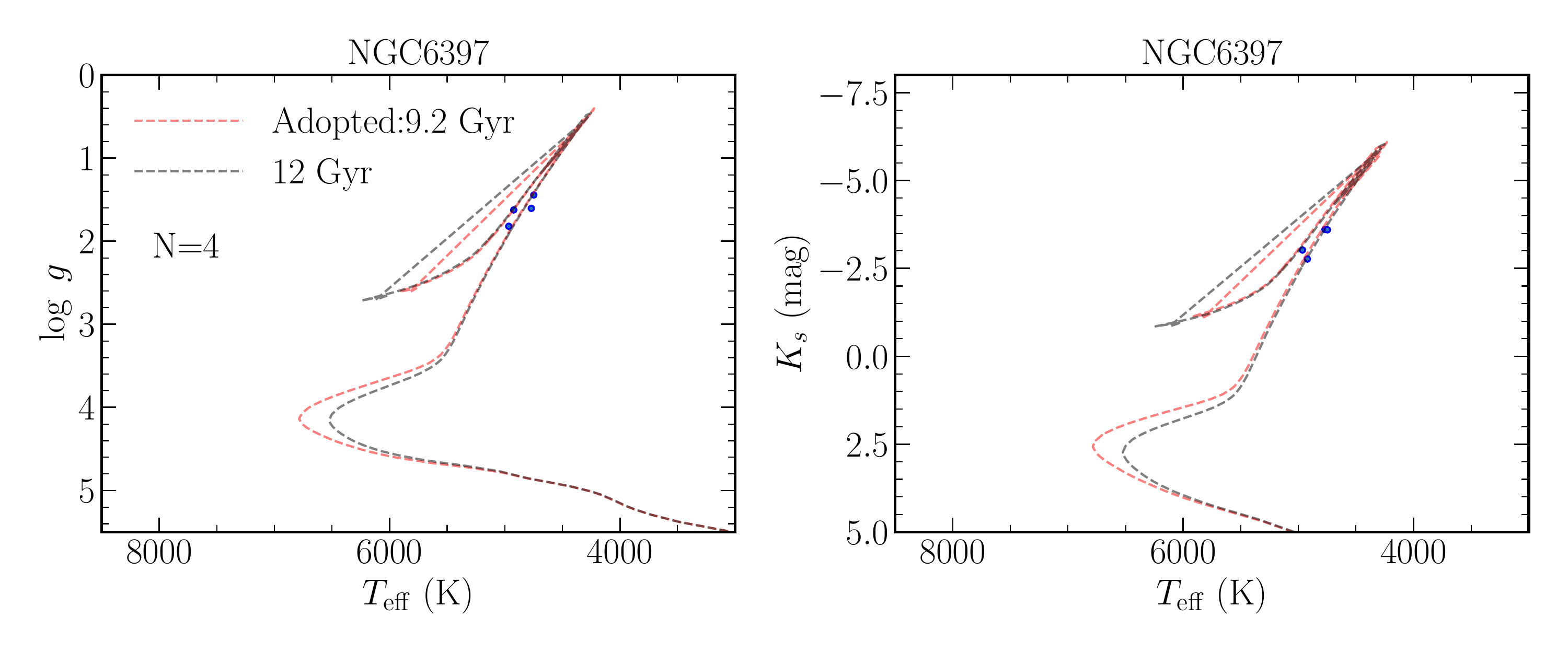}\\
\includegraphics[width=0.75\linewidth, angle=0]{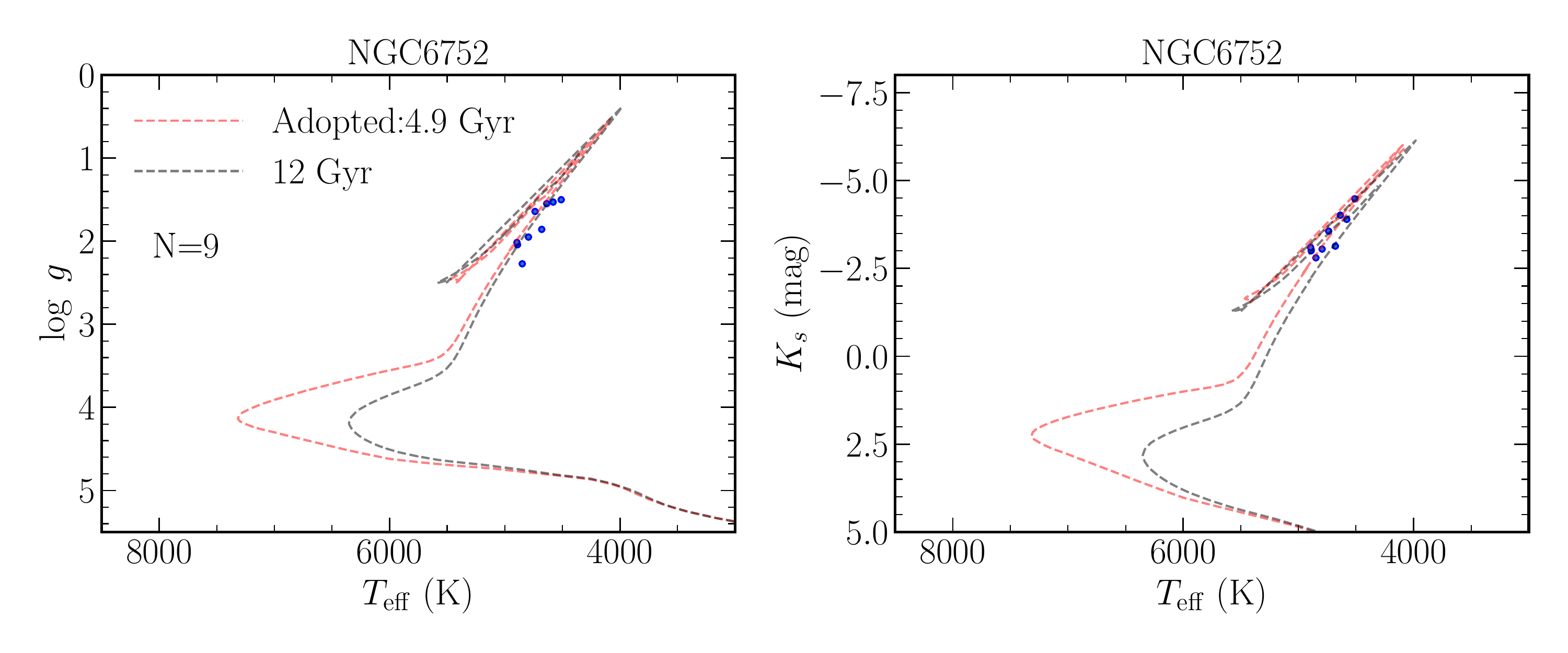}
\caption{\teff-\logg~(left) and \teff-$M_{\rm K_s}$ (right) diagrams for few cherry-picked globular clusters. %, which names are written at the top of each plot. 
Candidate stars have been selected based on the results of \citet{Gaia_Helmi18}. The isochrones with the adopted mean age (in red) and for 12\Gyr~are plotted in red and black, respectively, for the mean metallicity of the cluster as derived from the calibrated \gspspec~values.}
\label{fig:globular_cluster_comparisons_isochrones}
\end{center}
\end{figure}

Figures~\ref{fig:cluster_comparisons_isochrones} and \ref{fig:globular_cluster_comparisons_isochrones} show targets selected as being part of open and globular clusters, in the \teff-\logg~ space, as well as in the \teff-$M_{\rm K_s}$ space. In red are plotted the isochrones with metallicity and age equal to the mean \gspspec~(calibrated) metallicity and mean derived age of the cluster stars, whereas in black we plot the isochrones for the reference age (12\Gyr~in the case of the globular clusters). 

\end{appendix}

\end{document}